\documentclass[
    twocolumn,
    amsfonts,
    amsmath,
    amssymb,
    nobibnotes,
    aps,
    prd,
    nofootinbib,
    superscriptaddress,
]{revtex4-2}

\usepackage{amsmath}
\usepackage{amsfonts}
\usepackage{amssymb}
\usepackage{booktabs}
\usepackage[T1]{fontenc}
\usepackage{graphicx}
\usepackage[colorlinks=true, linkcolor=blue, urlcolor=blue, citecolor=blue]{hyperref}
\usepackage{siunitx}
\usepackage{xcolor}

\usepackage{graphicx}	
\usepackage{amsmath}	
\usepackage{amssymb}	
\usepackage{siunitx}
\usepackage{xcolor}
\usepackage{orcidlink}

\DeclareSIUnit\year{yr}
\DeclareSIUnit\parsec{pc}

\newcommand{\crit}{z}
\newcommand{\depth}{\mathcal{D}}
\newcommand{\fasttracks}{\texttt{fasttracks}}
\newcommand{\hyp}[1]{\mathcal{H}_{\mathrm{#1}}}
\newcommand{\pdet}{p_{\mathrm{det}}}

\newcommand{\Nsft}{N_{\mathrm{SFT}}}
\newcommand{\Sn}{S_{\mathrm{n}}}
\newcommand{\Tsft}{T_{\mathrm{SFT}}}

\newcommand{\A}{\mathcal{A}}
\newcommand{\hA}{\hat{\A}}
\newcommand{\unif}{\xi}

\newcommand{\Ntemplates}{N_{\mathrm{T}}}
\newcommand{\thresh}{\tau}

\newcommand{\UIB}{
Departament de F\'isica, Universitat de les Illes Balears, 
IAC3 -- IEEC, Carretera Valldemossa km 7.5, E-07122 Palma, Spain
}
\newcommand{\milan}{Dipartimento di Fisica ``G. Occhialini'', 
Universit\`a degli Studi di Milano-Bicocca, Piazza della Scienza 3, 20126 Milano, Italy}
\newcommand{\infn}{INFN, Sezione di Milano-Bicocca, 
Piazza della Scienza 3, 20126 Milano, Italy}

\begin{document}
\title{
    One-stop strategy to search for long-duration gravitational-wave signals
}

\author{Rodrigo Tenorio\,\orcidlink{0000-0002-3582-2587}}
\email{rodrigo.tenorio@unimib.it}
\affiliation{\milan}
\affiliation{\infn}
\affiliation{\UIB}

\author{Joan-René Mérou\,\orcidlink{0000-0002-5776-6643}}
\affiliation{\UIB}

\author{Alicia M. Sintes\,\orcidlink{0000-0001-9050-7515}}
\affiliation{\UIB}

\begin{abstract}
    Blind continuous gravitational-wave (CWs) searches are a significant computational 
    challenge due to their long duration and weak amplitude of the involved signals. 
    To cope with such problem, the community has developed a variety of data-analysis strategies
    which are usually tailored to specific CW searches; this prevents their applicability
    across the nowadays broad landscape of potential CW source. Also, their sensitivity
    is typically hard to model, and thus usually requires a significant computing investment.
    We present \fasttracks{}, a massively-parallel engine to evaluate detection statistics for 
    generic CW signals using GPU computing. We demonstrate a significant increase in computational
    efficiency by parallelizing the brute-force evaluation of  detection statistics without using
    any computational approximations. Also, we introduce a simple and scalable post processing
    which allows us to formulate a generic semi-analytic sensitivity estimate algorithm.
    These proposals are tested in a minimal all-sky search in data from the third observing run 
    of the LIGO-Virgo-KAGRA Collaboration. The strategies here discussed will become increasingly 
    relevant in the coming years as long-duration signals become a standard observation of 
    future ground-based and space-borne detectors.
\end{abstract}

\maketitle

\section{Introduction}

Continuous gravitational waves (CWs) are long-duration gravitational-wave 
signals (GWs)~\cite{Riles:2022wwz}. For the current network of detectors
operated by the LIGO-Virgo-KAGRA Collaboration 
(LVK)~\cite{LIGOScientific:2014pky,VIRGO:2014yos,KAGRA:2018plz},
the expected source of these signals are rapidly-rotating non-axisymmetric
neutron stars in our Galaxy~\cite{Sieniawska:2019hmd,Piccinni:2022vsd,2023NatAs...7.1160H}, 
although more exotic scenarios such as the evaporation of boson clouds around spinning black 
holes~\cite{Isi:2018pzk,Sun:2019mqb,Palomba:2019vxe, Zhu:2020tht,Jones:2023fzz}, dark matter 
halos~\cite{Miller:2022wxu,Miller:2022wxu,Miller:2023kkd,KAGRA:2024ipf, Bhattacharya:2024pmp},
or sub-solar mass compact binaries~\cite{Guo:2022sdd,Miller:2024khl,Miller:2024fpo} 
have been proposed as plausible sources, enriching the scientific scope of CW searches.
Due to their extended duration, these signals are modulated by the motion of 
the detectors~\cite{Jaranowski:1998qm, Cutler:2005hc}, and span a parameter space at the edge of 
current computing capabilities~\cite{Brady:1998nj, Wette:2014tca}.

Future ground-based and space-borne detectors (Einstein Telescope~\cite{Maggiore:2019uih},
Cosmic Explorer~\cite{Reitze:2019iox}, LISA~\cite{2017arXiv170200786A}, Taiji~\cite{Hu:2017mde},
TianQin~\cite{Li:2024rnk}) will extend the operational bandwidth into lower frequencies
of the GW spectrum. In such regime, GWs from compact binary coalescences (including those 
detected today using LVK data~\cite{KAGRA:2021vkt, Nitz:2021zwj, Wadekar:2023gea}) are expected to be observable 
for a significantly longer time~\cite{LIGOScientific:2016wyt}, bringing them qualitatively closer to CWs.
This suggests current GW data-analysis algorithms will have to be revisited 
to make an efficient use of the available computing resources.

These past decades have seen a wide variety of developments to search for CWs from unknown sources
(also known as \emph{blind searches})~\cite{Jaranowski:1998qm,Astone:2000jz,Krishnan:2004sv,Wette:2023dom}.
Fundamentally, most of these approaches are distinguished by their specific sensitivity-cost trade-offs,
which materialize in the form of different detection statistics, post-processing methods,
and follow-up steps~\cite{Tenorio:2021wmz}.

Although very successful, implementation details prevent their general
applicability to more general phenomena, such as NS glitches~\cite{Ashton:2017wui, Ashton:2018qth},
spin wandering~\cite{Mukherjee:2017qme}, proper motion~\cite{Covas:2020hcy},
parallax effects~\cite{Sieniawska:2022bcn}, post-glitch~\cite{Prix:2011qv}
and post-merger~\cite{msMagnetar,Sarin:2018vsi} CW-like signals,
or different kinds of bosons~\cite{Siemonsen:2022yyf}. Similar problems will be faced by 
compact-binary-coalescence analyses as we progress into next-generation
detectors~\cite{Miller:2023rnn}. These science cases, which are relevant for both the
first detection of a CW signal and future GW searches in general, require a general and efficient
long-duration GW search pipeline in order to understand the current state of the art and 
inform future developments in GW data analysis.

We present \fasttracks{}, a massively-parallel engine to compute short-coherence 
detection statistics for long-duration GW searches. \fasttracks{} 
parallelizes the evaluation of detection statistics based on \emph{time-frequency tracks}
across multiple templates; in particular, it is capable of implementing any detection statistic 
based on Short Fourier Transforms (SFTs)~\cite{sft,Astone_2005,Piccinni:2018akm}. 
Our results show that the use of GPU-accelerated brute-force
template evaluation provides comparable computing efficiencies to using model-specific optimizations
such as~\cite{Krishnan:2004sv, 2014PhRvD..90d2002A, Miller:2018rbg, Oliver:2019ksl, Covas:2019jqa}.
We release \fasttracks{} as an open-source Python package~\cite{fasttracks}.

Building on \fasttracks{} we propose a simple post-processing strategy to
efficiently parallelize the identification of interesting candidates. With this,
We generalize the semi-analytic sensitivity estimate algorithms introduced 
in~\cite{Wette:2011eu, Dreissigacker:2018afk} to a broader class of detection
statistic and to account for the effect of post-processing steps on the search.
This method, which we released under the open-source~\texttt{cows3} 
Python package~\cite{Mirasola:2024lcq,cows3}, will significantly reduce the computing cost
associated to estimating the sensitivity of blind CW searches.

We test our proposals on a blind search for binary CW sources 
using data from the LVK's third observing run~\cite{KAGRA:2023pio}.

The paper is structured as follows: We review
short-coherence detection statistics in Sec.~\ref{sec:statistics} and present an efficient
implementation using GPUs in Sec.~\ref{sec:gpu_stats}. In Sec.~\ref{sec:postprocessing} we discuss
a simple template bank and post-processing strategy. These results are collected in Sec.~\ref{sec:sensitivity_estimate}
to propose a generalized semi-analytic sensitivity estimate method. We test the performance
of our proposals on real data in Sec.~\ref{sec:real_data}. We conclude in Sec.~\ref{sec:conclusion}.

\section{Short-coherence CW searches\label{sec:statistics}}

Blind CW searches use \emph{semicoherent} methods in order to cover broad parameter-space regions
under affordable computing budgets~\cite{Krishnan:2004sv, Cutler:2005pn, Prix:2012yu}. The overall idea
of a semicoherent method is to divide data into segments that are analyzed independently to then construct
a combined detection statistic. Different semicoherent methods are possible,
depending on the chosen duration of the data segments and how different detectors are combined.

In this work we will treat the case of ``short-coherence'' CW searches, in which 1) different detectors are
treated independently and 2) CW signals are monochromatic within the duration of a data segments. 
Simply put, a short-coherence detection statistic combines the Fourier power contained in SFTs of the data
(see Fig.~\ref{fig:spectrogram}), as opposed to fully-coherent statistics which combine complex Fourier amplitudes.
This denomination includes most of the current searches based on normalized-power or number-count 
statistics~\cite{Krishnan:2004sv, powerflux, powerflux2, 2014PhRvD..90d2002A, Miller:2018rbg, Oliver:2019ksl,
Covas:2019jqa}. The use of a short coherence time increases their robustness to unmodeled physics 
such as NS glitches or spin wandering~\cite{Ashton:2017wui, Mukherjee:2017qme}.

This section summarizes the main results in short-coherence detection statistics.
The results here presented will be referred to later in the paper whenever new developments are presented.

\subsection{Signal model}

We use the standard quasi-monochromatic CW signal model~\cite{Wette:2023dom} parametrized by two sets of parameters:
the amplitude parameters $\A$ and the frequency-evolution parameters $\lambda$. The amplitude parameters
include the CW amplitude $h_0$, the (cosine of the) inclination angle $\cos\iota$, the polarization angle $\psi$,
and the initial phase $\phi_0$. The frequency-evolution parameters $\lambda$ depend on the signal of interest. 
The split is motivated by the structure of a CW signal as detected by an interferometric
detector~\cite{Jaranowski:1998qm}:
\begin{equation}
    h(t) = \sum_{\mu = 0}^{3} \A_{\mu} h_{\mu}(t; \lambda)  \,,
\end{equation}
where the four coefficients $\A_{\mu}$ depend solely on $\A$. This is a direct consequence the GW lasting for longer
than a day, causing detector-induced modulation to become relevant~\cite{Covas:2022xyd,Tenorio:2025yca}.
We define $\hA = \{ \cos \iota, \psi\}$ for later convenience.
Individual components of $\lambda$ will be indexed using square brackets $\lambda[k]$.

The results in this work will be translatable to any quasi-monochromatic GW signal; that is,
any signal with a narrow Fourier spectrum in a sufficiently short time-scale
(see Ref.~\cite{Tenorio:2025yca} for other applications of this approach).
Specific numerical results will be given for the case of a CW source in a circular binary
system~\cite{Messenger:2011rg,Leaci:2015bka,Covas:2019jqa},
\begin{equation}
    f(t; \lambda) = f_0 
    \left[1 
    + \frac{\vec{v}(t)}{c} \cdot \hat{n} 
    - a_{\mathrm{p}} \Omega \cos{\left(\Omega t - \phi_\mathrm{b} \right)} 
    \right]
    \label{eq:track}
\end{equation}
i.e. $\lambda = \{ f_0, \hat{n}, a_{\mathrm{p}}, \Omega, \phi_{\mathrm{b}}\}$,
where $f_0$ is the CW frequency at the start of the observation in the Solar System Barycenter, 
$\hat{n}$ is the sky position of the source, $a_{\mathrm{p}}$ is the projected
semimajor axis, $\Omega$ is the binary orbital frequency, and \mbox{$\phi_{\mathrm{b}} =  t_{\mathrm{asc}} \Omega$}
is an orbital phase where $t_{\mathrm{asc}}$ is the time of passage through the ascending node.
Here, the binary orbital parameters $a_{\mathrm{p}}, \Omega, \phi_{\mathrm{b}}$ account for the Doppler modulation
suffered by the CW signal due to the motion of the source in the binary system. For $a_{\mathrm{p}}=0$, the source
effectively behaves like an isolated one. Note that the source's intrinsic frequency evolution (i.e. spindown terms)
is neglected; this is a reasonable assumption for short-coherence searches aiming at sources in binary
systems~\cite{Covas:2019jqa}.

\subsection{Short-coherence detection statistics}

We assume $x$ consists of a series of Short Fourier Transforms~\cite{Krishnan:2004sv,sft}
with duration $\Tsft$  labeled by an index $\alpha$ and a detector-index $X$\footnote{
Consecutive timestamps may differ by more than $\Tsft$ if there are gaps in the data.}.
The value of an SFT at timestamp $t$ and frequency
$f$ will be denoted as $\tilde{x}[t; f]$. The total number of SFTs will be denoted as $\Nsft$.

The general form of a short-coherence detection statistic $s$ is given by
\begin{equation}
    s(\lambda) = \sum_{X, \alpha} w_{X\alpha}(\hA, \lambda) s[t_{X\alpha}; f(t_{X\alpha}; \lambda)] \,.
    \label{eq:short_coh_general}
\end{equation}
Here,  $w_{X\alpha}(\hA, \lambda)$ is an SFT-dependent weight encoding information about the noise floor,
the detector's sensitivity, and possibly the signal's amplitude parameters. Whenever required we will use
the weights introduced in~\cite{Palomba_2005}, which are suitable for all-sky searches for isotropically-oriented
sources~\cite{LIGOScientific:2007hnj}:
\begin{equation}
    w_{X\alpha}(\hat{n}) \propto 
    \frac{a^2(t_{X\alpha}; \hat{n}) + b^2(t_{X\alpha}; \hat{n})}{S_{\mathrm{n}X\alpha}} \,.
\end{equation}
$a$ and  $b$ are the detector's response functions~\cite{Jaranowski:1998qm} and 
$S_{\mathrm{n}X\alpha}$ is the averaged single-sided power spectral density (PSD)
of the noise in the frequency band under analysis, which for CW searches usually spans
about \SI{0.1}{\hertz}. Throughout this work, the PSD is estimated by means 
of a running-median estimate for each SFT $X\alpha$ and frequency $f$~\cite{prixCFSv2}; 
this dependency will be kept implicit in the notation, i.e. $\Sn = S_{\mathrm{n}X\alpha}[f]$. 
Other pipelines use different methods for this estimate, such as autoregressive processess~\cite{Astone_2005}.

\begin{figure}
    \includegraphics[width=\columnwidth]{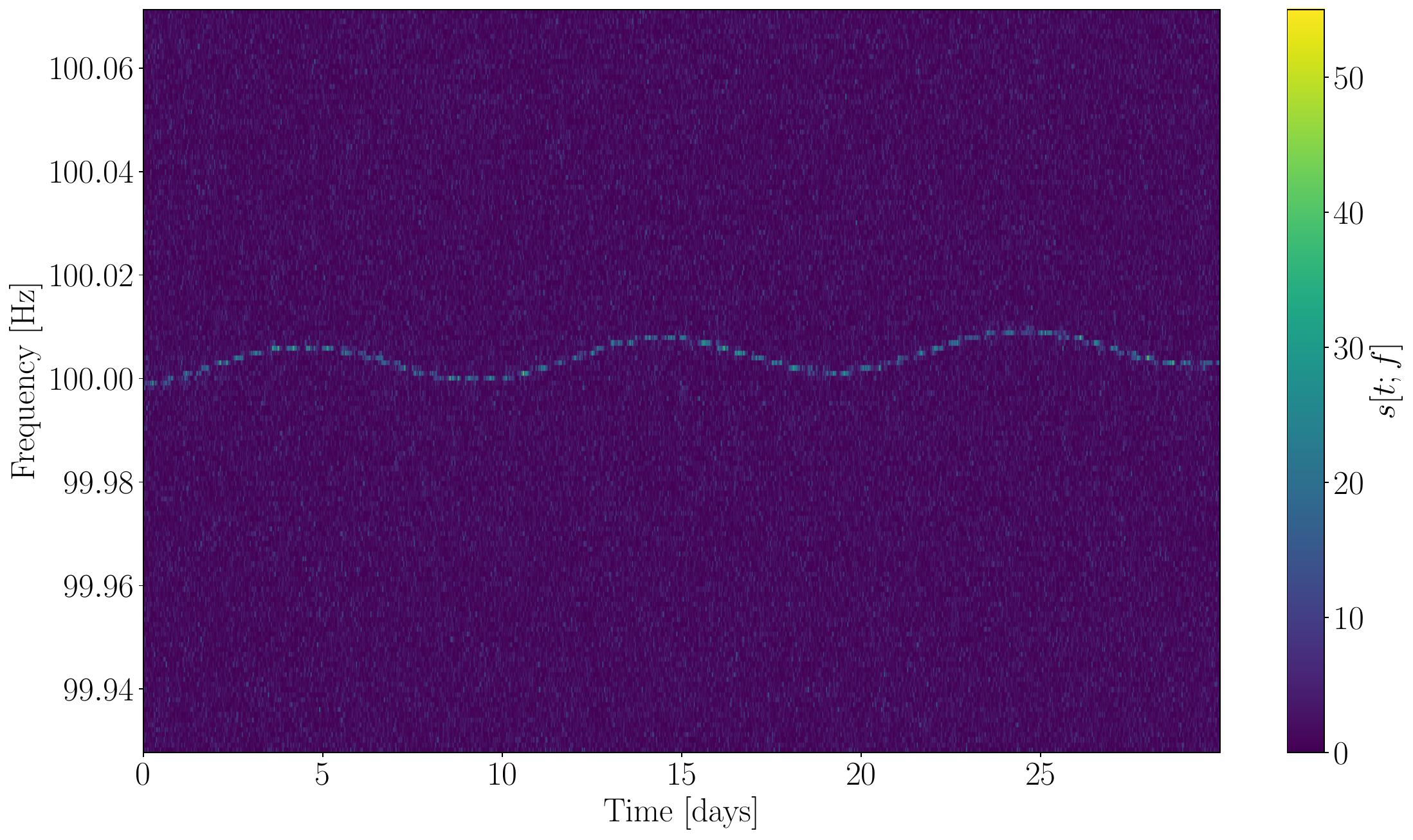}
    \caption{
        Normalized spectrogram of 1 month of narrow-band simulated Gaussian noise as measured by the LIGO Hanford detector
        using $\Tsft = \SI{1024}{\second}$. Data contains Gaussian noise with $\Sn = 10^{-23}$ 
        and a binary CW signal with an amplitude of $h_0 = 2 \times 10^{-24}$
        visible as a high-power oscillating track at about \SI{100}{\hertz}.  
        Amplitude modulations are due to the anisotropy of the detector,
        while frequency modulations are due to the binary orbital period of 10 days. 
        Daily and yearly Doppler modulations are too small and 
        too long, respectively, to be observed at these scales.
    }
    \label{fig:spectrogram}
\end{figure}

Weights are defined up to a constant which we are free to set so that
\begin{equation}
    \sum_{X, \alpha} w_{X\alpha}(\hat{n}) = 1 \,.
\end{equation}
$s[t; f]$, on the other hand, is a function of $\tilde{x}[t;f]$ at the given timestamp $t$ 
but possibly an arbitrary number of frequencies~\cite{Allen:2002bp}.
In this derivation we will focus on
\emph{normalized power}\footnote{Note that this corresponds to $2\rho_{k}$ in \cite{Krishnan:2004sv}.}
\begin{equation}
    s[t;f] 
    = \frac{4}{\Tsft \Sn} \left| \tilde{x}[t;f] \right|^2 \,.
    \label{eq:s_tf}
\end{equation}
The set of all $s[t; f]$ values is commonly referred to as the \emph{normalized spectrogram}
of the data, an example of which we show in Fig.~\ref{fig:spectrogram}. 
Equation~\eqref{eq:short_coh_general} could thus be intuitively described as the
integration of power following a time-frequency ``track'' defined by a 
specific template $\lambda$ across the spectrogram.

Additionally in Sec.~\ref{sec:real_data} we will make use of the \emph{number count}, which is defined
as~\cite{Krishnan:2004sv}
\begin{equation}
    \mathrm{nc}[t;f] =\left\{ \begin{matrix} 1 & \mathrm{if} \,  s[t; f]> 3.2 \\ 0 & \mathrm{otherwise} \end{matrix}\right\}\,.
    \label{eq:number_count}
\end{equation}
and can be used to define the \emph{weighted number count} akin to Eq.~\eqref{eq:short_coh_general}.
The threshold $3.2$ corresponds to the optimal choice in Gaussian noise, in the sense discussed in~\cite{Krishnan:2004sv};
other definitions are thoroughly used in the literature~\cite{2014PhRvD..90d2002A}.

We now derive the distribution of $s$ under both the noise and signal hypotheses.

\subsubsection{Distribution under the noise hypothesis\label{subsec:noise}}

Under the noise hypothesis, data is assumed to consist of Gaussian noise 
\begin{equation}
    \mathcal{H}_{\mathrm{G}}: \, \tilde{x} = \tilde{n} \,,
\end{equation}
where the real $\mathfrak{R}$ and imaginary $\mathfrak{I}$ parts follow the same distribution given by
\begin{equation}
    \mathfrak{R} \tilde{n}[t;f], \mathfrak{I} \tilde{n}[t;f] \sim \mathrm{Gauss}\left(0, \frac{1}{2} \sqrt{\Tsft \Sn}\right) \,,
\end{equation}
and the standard deviation follows from the Wiener-Kintchine theorem.

Normalized power is then given by
\begin{equation}
    \mathcal{H}_{\mathrm{G}}: \, s[t;f] =  \frac{4}{\Tsft \Sn}\left|\tilde{n}[t; f]\right|^2\,.
\end{equation}
Note that the variance of both $\mathfrak{R}\tilde{n}$ and $\mathfrak{I}\tilde{n}$ is equal and given by $\Tsft \Sn / 4$.
As a result, $s[t; f]$ is the sum of two zero-mean unit-variance Gaussian variables and thus follows a chi-squared distribution
with 2 degrees of freedom
\begin{equation}
    s[t;f] | \mathcal{H}_{\mathrm{G}} \sim \chi^2_{2} \,.
    \label{eq:s_chi2}
\end{equation}
Normalized power $s(\lambda)$ is thus the weighted sum of chi-squared distributed variables. This distribution does not have a
simple closed form, but it can be efficiently simulated by generating $\Nsft$ draws from a $\chi^2_2$ distribution and computing
the corresponding weighted sum. In the limit of $\Nsft \gg 1$, (typically \mbox{$\Nsft \sim \mathcal{O}(10^{3-4})$}), Lyapunov's Central
Limit Theorem implies
\begin{equation}
    s(\lambda) | \hyp{G} \sim \mathrm{Gauss}(\mu_{\mathrm{G}}, \sigma_{\mathrm{G}})
\end{equation}
where
\begin{align}
    \mu_{\mathrm{G}} = 2 \sum_{X, \alpha} w_{X\alpha}\,,\\
    \sigma^{2}_{\mathrm{G}} = 4 \sum_{X, \alpha} w^2_{X\alpha}\,.
\end{align}
Since weights $w_{X\alpha}$ are defined up to a constant, we are free to choose a fixed
value for either $\mu_{\mathrm{G}}$ or $\sigma^2_{\mathrm{G}}$.

As noted in~\cite{reinhardBias}, this derivation makes the assumption that
the PSD $\Sn$ is \emph{known}, which is used implicitly in Eq.~\eqref{eq:s_chi2} as
\begin{equation}
    x \sim \mathrm{Gauss}(0, \sigma) \rightarrow x/\sigma \sim \mathrm{Gauss}(0, 1) \,.
    \label{eq:known_sigma}
\end{equation}
Since $\Sn$ is \emph{estimated} from the data, the estimated PSD becomes a random variable itself
and Eq.~\eqref{eq:known_sigma} ceases to be applicable. This implies that, strictly speaking, $s[t; f]$ should be described
using \emph{ratio distributions}, in a similar manner to~\cite{Rover:2008yp, Rover:2011qd, Sasli:2023mxr}. Our specific
implementation of PSD estimation follows that of~\cite{Krishnan:2004sv}, which uses a running-median estimate as implemented
in \texttt{XLALNormalizeSFT}~\cite{lalsuite}; as discussed 
in~\cite{reinhardBias} and corroborated by our numerical experiments, in such case the effect of estimating $\Sn$ can
be described by introducing a small upward bias in $\mu_{\mathrm{G}}$ and $\sigma_{\mathrm{G}}$. For our case,
the running median is computed using 101 frequency bins implying a bias of about $1.2\%$ for $\mu_{\mathrm{G}}$. 

\subsubsection{Distribution under the signal hypothesis}

Under the signal hypothesis data consist of Gaussian noise
and a CW signal $h$ with parameters $\lambda, \A$:
\begin{equation}
    \mathcal{H}_{\mathrm{S}}:\, \tilde{x} = \tilde{n} + \tilde{h}\,.
\end{equation}
As previously discussed, within an SFT $X\alpha$ a CW signal can be described as a monochromatic signal with
frequency $f(t_{X\alpha};\lambda)$ and amplitude given by~\cite{Krishnan:2004sv}
\begin{equation}
    \begin{aligned}
        & \left|\tilde{h}_{X\alpha}\right| \approx 0.7737  \\
        & \times \frac{\Tsft}{2} h_0 
    \left| F_{+}(t_{X\alpha}; \hat{n}, \psi) \frac{1 + \cos^2{\iota}}{2} +F_{\times}(t_{X\alpha}; \hat{n}, \psi) \cos\iota \right| \,
    \end{aligned}
\end{equation}
where $F_{+, \times}$ are the polarization-dependent antenna-pattern functions and 
the numerical $\approx 0.7737$ factor is due to the fact that the CW's frequency
in an SFT can be uniformly distributed around a Fourier bin~\cite{Allen:2002bp}.

The presence of a deterministic signal in a frequency bin shifts the noise's Gaussian distribution
without altering it's variance. Specifically (and respectively for the imaginary part)
\begin{equation}
    \mathcal{H}_{\mathrm{S}}:\, \mathfrak{R} \tilde{x}_{X\alpha}
    \sim \mathrm{Gauss}\left(\mathfrak{R}\tilde{h}_{X\alpha}, \frac{1}{2}\sqrt{\Tsft \Sn} \right)\,.
\end{equation}
This implies $s[t, f]$ is no longer zero-mean, but instead
\begin{equation}
    \rho^2_{X\alpha}
    = \langle s[t_{X\alpha}; f(t_{X\alpha};\lambda)] \rangle 
    = \frac{4}{\Tsft \Sn} \left|\tilde{h}(t_{X\alpha}; f(t_{X\alpha};\lambda))\right|^2  \,.
\end{equation}
As a result, $s[t, f]$ follows a non-central chi-squared distribution with two degrees of freedom
\begin{equation}
    \mathcal{H}_{\mathrm{S}}: \, s_{X\alpha} \sim  \chi^2_2\left( \rho^2_{X\alpha}\right) \,.
\end{equation}
Once more, the distribution of $s(\lambda) | \hyp{\mathrm{S}}$ does not have a closed form but can be efficiently
simulated by simulating the individual $s_{X\alpha}$ values. In the $\Nsft \gg 1$ limit Lyapunov's Central Limit
Theorem implies
\begin{equation}
    s(\lambda) | \hyp{\mathrm{S}}  \sim \mathrm{Gauss}(\mu_{\mathrm{S}}, \sigma_{\mathrm{S}})
    \label{eq:s_gauss}
\end{equation}
where
\begin{align}
    \mu_{\mathrm{S}}& 
    = \mu_{\mathrm{G}}  + \sum_{X,\alpha}  \rho^2_{X\alpha} w_{X\alpha} = \mu_{\mathrm{G}} + \rho^2_{1}\,\\
    \sigma^2_{\mathrm{S}} & 
    = \sigma^2_{\mathrm{G}} + \sum_{X,\alpha} \rho^2_{X\alpha} w^2_{X\alpha} = \sigma_{\mathrm{G}}^2 + \rho^2_{2} \,.
\end{align}

In this derivation we implicitly assumed we had access to the signal's frequency-evolution parameters $\lambda$.
In a search, however, we will cover the parameter space with a discrete set of templates,
implying that the \emph{closest} template to the signal, $\lambda'$, will recover a fraction of the fully-matched detection 
statistic. This fraction can be characterized by means of the mismatch~\cite{Prix:2006wm,Prix:2007ks,Allen:2021yuy}
\begin{equation}
        m(\lambda';\lambda) = \frac{s(\lambda) - s(\lambda')}{s(\lambda) - \mu_{G}}\,.
        \label{eq:mismatch}
\end{equation}
The effects of mismatch can be simply introduced by reducing the value of the non-centrality 
parameters $\rho^2_{1, 2} \rightarrow (1 - m) \rho^2_{1, 2}$ . The expected distribution of $m$ depends on the chosen template bank. 
We will focus on the case of random template banks in Sec.~\ref{sec:postprocessing}.

Note that, for a given observing run, the distribution of $s(\lambda)$ depends ultimately on four parameters
\mbox{$\{ \mu_{\mathrm{G}}, \sigma_{\mathrm{G}}, \rho^2_{1}, \rho^2_{2}\}$}. This will be exploited
in Sec.~\ref{sec:sensitivity_estimate} to efficiently estimate the sensitivity of a short-coherence CW search.
It will also be convenient to define the amplitude-independent quantity
\begin{equation}
    \hat{\rho}^2_{X\alpha}(\hA, \hat{n})= \left( \frac{\Sn}{h_0^2}\right) \rho^2_{X\alpha}(\hA, \hat{n}) \,,
\end{equation}
where the orientation angles in $\hA$ account for the source's orientation, while the sky position $\hat{n}$ accounts for the detector's sensitivity to that specific direction.

\section{Accelerating short-coherence detection statistics\label{sec:gpu_stats}}

The main stage of a blind CW search is to evaluate a detection statistic $s$ on a template bank,
which is a set of waveform templates $\{ \lambda_{i}, i = 1, \dots, \Ntemplates\}$ 
where each template $\lambda_{i}$ contains the relevant parameters to describe a signal and
usually $\Ntemplates \sim \mathcal{O}(10^{12-18})$~\cite{Wette:2023dom}.
Much of the CW literature is devoted to  design smart search pipelines that exploit 
parameter-space correlations to make the computation of $s$ more  
efficient, usually with a small cost in sensitivity~\cite{Williams:1999nt,Krishnan:2004sv,Pletsch:2009uu,
Patel:2009qe,Dergachev:2010tm, Dergachev:2011pd, powerflux, powerflux2,2014PhRvD..90d2002A,
Meadors:2017pef,Wette:2018bhc,Miller:2018rbg, Dergachev:2018ftg,Oliver:2019ksl,Dergachev:2019wqa,
prixCFSv2,Covas:2019jqa,Covas:2022mqo,Andres-Carcasona:2023zny,Miller:2024jpo}.

Lately, several of these pipelines have been \emph{re-implemented} using \texttt{CUDA}~\cite{cuda}
(or similar low-level languages) to benefit from GPU acceleration~\cite{Keitel:2018pxz,Covas:2019jqa,
Rosa:2021ptb,Dunn:2022gai}. Specifically, these works re-implement a pre-existing computationally
efficient strategy, such as \emph{Partial Hough Maps}~\cite{Krishnan:2004sv} or 
the $\mathcal{F}$-statistic's \emph{Resampling} 
algorithm~\cite{Patel:2009qe,Meadors:2017pef,polgraw}, 
and parallelize its evaluation on a GPU. This results in the \emph{batch-evaluation} of
waveform templates across limited portions of the parameter space (e.g. multiple sky 
positions~\cite{Krishnan:2004sv} for a subset of frequency bins or multiple frequency bins
for a fixed sky position~\cite{Patel:2009qe, Meadors:2017pef, polgraw}. 

These approaches, while effective, were developed with a different computational landscape in
mind, and may not be suitable (or required) given the computing capabilities of
contemporary hardware. Also, as previously discussed, they may prevent their generalization 
to other astrophysically-plausible sources.

Here, we parallelize the brute-force evaluation of a short-coherence detection statistics without
using pipeline-specific implementation tricks.
We implement the computation of Eq.~\eqref{eq:short_coh_general} in \fasttracks{}~\cite{fasttracks} 
using \texttt{jax}~\cite{jax, jax2018github}, which for our purposes can be thought as a
GPU-capable implementation
of Python's array library \texttt{numpy}~\cite{harris2020array}.
Any other GPU-capable array library would be equally feasible
for this implementation~\cite{cupy_learningsys2017, 2019arXiv191201703P}. 
The use of such high-level languages, which allow for a fast development cycle and 
simple code maintenance, is in stark contrast with other GPU-capable 
CW-searches~\cite{Keitel:2018pxz,Covas:2019jqa,Dunn:2022gai}.

A brute-force single-template evaluation of $s(\lambda)$ uses the frequency-evolution track $f(t; \lambda)$ 
to sum individual spectrogram bins $s[t;f(t; \lambda)]$. These operations are trivially implemented using array slicing 
and reduce operations.  To evaluate multiple templates at once, we \emph{vectorize} $s(\lambda)$ using \texttt{jax}'s \texttt{vmap}
function\footnote{
    We could also vectorize by hand using other array frameworks~\cite{cupy_learningsys2017, 2019arXiv191201703P},
    but we choose to use \texttt{vmap} for its conceptual simplicity.
}. More precisely, we denote a \emph{batch} of $n$ templates  by $\vec{\lambda} = \{ \lambda_{1}, \dots, \lambda_{n} \}$.
\texttt{vmap} promotes a single-template function $s: \lambda \rightarrow s(\lambda)$ to 
a vectorized function \mbox{$\texttt{vmap} \, s : \vec{\lambda} \rightarrow \{s(\lambda_i), \, i = 1, \dots, n \}$} in such
a way that the evaluation of different templates is parallelized on the available computing device. 
Note that we made \emph{no assumption} on the properties of $\vec{\lambda}$
(i.e. we are not restricted to fixed frequencies or fixed sky positions).

\begin{figure}
    \includegraphics[width=\columnwidth]{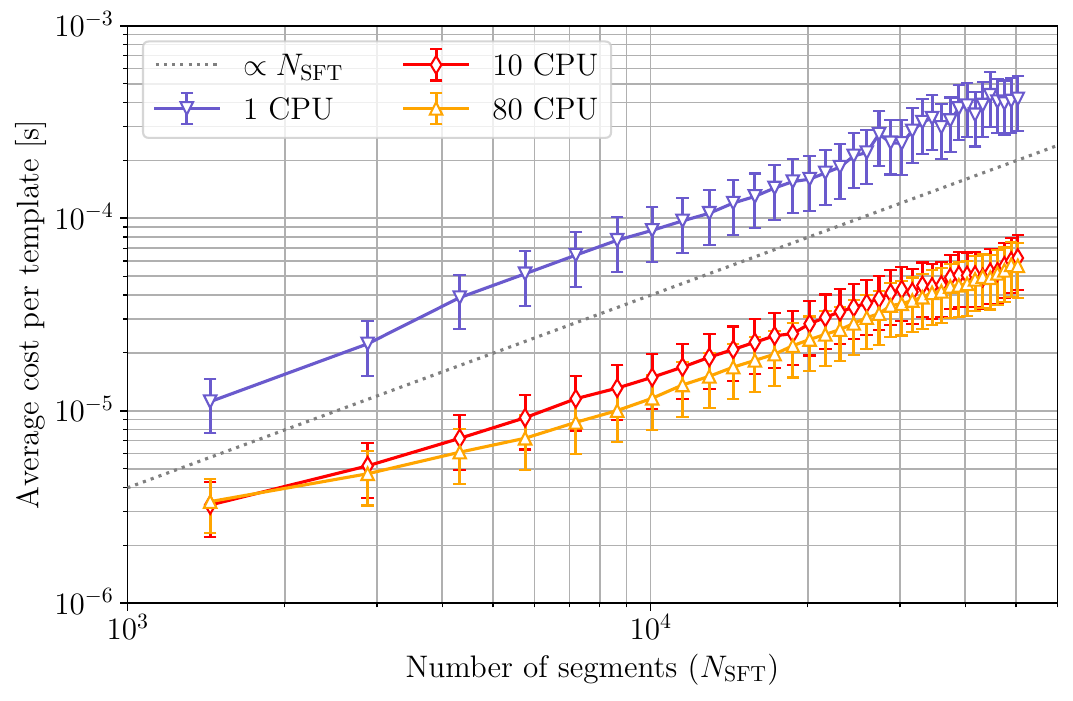}
    \includegraphics[width=\columnwidth]{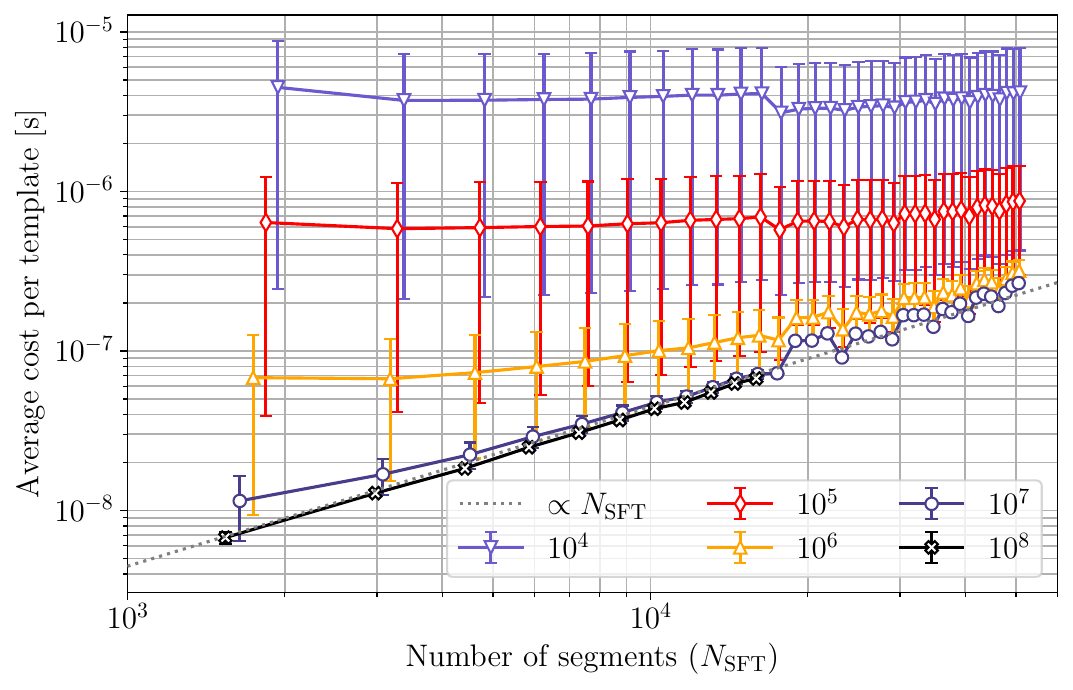}
    \caption{
        Cost of evaluating $\mathtt{vmap}\, s$ on 
        (upper panel) a single computing node with multiple
        Intel Xeon Platinum 8460Y CPUs with a fixed batch size of $10^4$ and
        (lower panel) a single NVIDIA Hopper H100 GPU with multiple batch sizes.
        The vertical axis shows the average evaluation cost of a template,
        computed by evaluating a batch and dividing by the number of elements in the batch.
        The horizontal axis shows the number of SFTs in the dataset.
        The dotted line is proportional to the number of SFTs and represents the point at
        which the computation becomes dominated by the summation of powers.
        Each measurement is the average across 10 different realizations. 
        Points with the same $\Nsft$ are nudged to avoid overlap of markers.
        }
    \label{fig:timings_gpu}
\end{figure}

We benchmark the GPU-accelerated evaluation of $\texttt{vmap}\,s$ [Eq.~\eqref{eq:short_coh_general}] on an NVDIA Hopper
H100 GPU using a random  template-bank of templates consistent with a CW source in a binary system.
Templates are randomly sampled in frequency, sky position, and binary orbital parameters as discussed
in Sec.~\ref{sec:postprocessing}. For comparison, we repeat the same benchmark for a fixed batch size of 
$10^4$ templates parallelizing the evaluation over multiple CPU cores using a CPU Intel Xeon Platinum 8460Y.
This batch size is so that the CPU computing capability is saturated.
The resulting average cost per template is shown in Fig.~\ref{fig:timings_gpu}.

We find the GPU implementation to be two to three orders of magnitude faster than the CPU implementation.
These results are comparable to GPU implementations of \emph{specific} search pipelines such
as~\cite{Covas:2019jqa, Rosa:2021ptb} using similar statistics and number of SFTs.
We thus conclude that \fasttracks{}~\cite{fasttracks} provides a simple,
efficient, and easily-generalizable implementation of short-coherence detection statistics. 

We find that the average computing cost per template on a GPU $c_{\lambda}$ is described by the following model
\begin{equation}
    \log_{10} c_{\lambda} = 
    -\frac{5}{3} \log_{10} n 
    +\frac{\log_{10} \Nsft}{1 + \left(10^6 /n\right)^{5/6}} 
\end{equation}
This includes generating random templates, computing the corresponding frequency indices, 
gathering them, and then reducing the operation. The computational efficiency of \fasttracks{}
is primarily  dominated by the number of templates evaluated in parallel.
Upon saturating the capacity of the GPU, however, the cost becomes dominated by the number of $\Nsft$,
which is proportional to the amount of numbers to add up to compute a statistic
(i.e. the average cost approaches the dotted line in Fig.~\ref{fig:timings_gpu}).
This was already noted in~\cite{Covas:2022mqo}, where the main semicoherent computation is run
over a smaller amount of coherent segments rather than over SFTs.
Further computing gains are thus expected from the summation of less semicoherent segments.

These results show that 
\emph{significant speed ups can be achieved solely due to the batch-evaluation of a 
high number of  waveform templates} on a GPU, \emph{regardless of whether a pipeline-specific
optimization is being used}. 
This automatically generalizes  the results discussed 
in~\cite{Keitel:2018pxz,Covas:2019jqa, Rosa:2021ptb,Dunn:2022gai} to \emph{any} 
sort of short-coherence search as long as a frequency-evolution model $f(t;\lambda)$ is available.

\section{Compressing wide parameter-space searches\label{sec:postprocessing}}

As discussed in Sec.~\ref{sec:gpu_stats}, blind CW searches analyze a formidable amount
of templates to cover the parameter space under consideration. Templates are usually 
deployed using lattices~\cite{Prix:2009tq, Wette:2014tca,Wagner:2021hgv,Allen:2021yuy} to 
benefit from computing optimizations~\cite{Krishnan:2004sv, 2014PhRvD..90d2002A, 
Pletsch:2009uu, Pletsch:2010xb, Wette:2018bhc}.  Upon completion, the sensitivity of a 
search is generally dominated by the ``loudest candidates'' (i.e. templates with the maximum 
detection statistic)~\cite{Keitel:2019zhb, Tenorio:2021wad,  LIGOScientific:2021quq, Wette:2021tbv}. 
Search sensitivity can be increased by following-up an increasing number of candidates, 
especially if the parameter space is polluted by instrumental artifacts~\cite{2018PhRvD..97h2002C, 
Mirasola:2024lcq}.
This motivates the use of post-processing methods~\cite{Singh:2017kss,
Morawski_2020,Tenorio:2020cqm,Beheshtipour:2020zhb,Beheshtipour:2020nko,Steltner:2022aze},
to ascribe candidates to common causes. As a result, correlated candidates are analyzed together
and the sensitivity increases, for a given number of selected candidates, compared analyzing individual candidates.

The computational gains demonstrated in Sec.~\ref{sec:gpu_stats} grant us the freedom to rethink the 
template-bank (Sec.~\ref{subsec:uniform_tb}) and post-processing (Sec.~\ref{subsec:aot_clustering}) setups
to accommodate our needs. We aim to formulate a simple, easy-to-model signal-agnostic setup that fully exploits
the capabilities of \fasttracks{}. We will make use of these developments in Sec.~\ref{sec:sensitivity_estimate}
to provide a generic sensitivity-estimate method and in Sec.~\ref{sec:real_data} to deploy a search in real data.

\subsection{Template bank setup\label{subsec:uniform_tb}}

Two main ingredients are required to set a template bank up, namely a notion of ``closeness'' amongst templates and
a template-placement prescription. These have been thoroughly studied for long-coherence 
statistics~\cite{Brady:1998nj, Astone:2000jz, Pletsch:2009uu, Pletsch:2010xb, Messenger:2011rg,Wette:2013wza,
Wette:2014tca, Leaci:2015bka, Wette:2016raf, Wette:2018bhc,Wagner:2021hgv},
but not so much for short-coherent searches (with the notable exception of~\cite{Covas:2022mqo}).
We here propose a simple approach to construct template banks for short-coherence detection statistics.
While our specific application will be a blind search for CW sources in binary systems, the general principles
here described should be applicable to most kinds of CW searches.

\subsubsection{A notion of ``closeness''}

Short-coherence detection statistics [Eq.~\eqref{eq:short_coh_general}] depend on $\lambda$ mainly through the 
frequency-evolution model $f(t;\lambda)$, from which a parameter-space distance can be defined~\cite{Tenorio:2020cqm}.
For a given spectrogram time-resolution $\Tsft$, frequency is naturally discretised in steps of $\Tsft^{-1}$. We can thus
define the parameter-space resolution of a given parameter $\delta \lambda[k]$ as the minimal variation so that
the resulting frequency track differs by about one frequency bin along the frequency evolution:
\begin{equation}
    \delta \lambda[k] = \frac{1}{\Tsft} \left| \frac{\partial}{\partial \lambda[k]} f(t;\lambda)\right|^{-1}\,.
    \label{eq:resolution}
\end{equation}
This prescription, which is standard for short-coherence detection statistics~\cite{Krishnan:2004sv, 2014PhRvD..90d2002A,
Oliver:2019ksl, Miller:2018rbg}, may yield time-dependent parameter-space resolutions, in which case we will conservatively
take the  minimum value of $\delta \lambda[k]$. As shown in Appendix~\ref{app:binary}, Eq.~\eqref{eq:resolution} yields 
the expected functional dependency of $\delta \lambda$ on $\lambda$ for the case of short-segment semicoherent
detection statistics~\cite{Krishnan:2004sv,Messenger:2011rg,Leaci:2015bka}.

For example, the resolutions for the orbital binary parameters are (see Appendix~\ref{app:binary})
\begin{equation}
    \begin{aligned}
    \delta a_\mathrm{p} = &\Tsft^{-1} \left(f_0 \Omega \right)^{-1} \,, \\
    \delta \phi_{\mathrm{b}} =&  \Tsft^{-1}  \left(f_0 a_{\mathrm{p}}\Omega\right)^{-1} \,, \\
    \delta \Omega =& \left(\Tsft T_{\mathrm{obs}}\right)^{-1} \left(f_0 a_{\mathrm{p}} \Omega \right)^{-1} \,,
    \end{aligned}
    \label{eq:example}
\end{equation}
in agreement with the resolutions derived from Eq.~(62) in Ref.~\cite{Leaci:2015bka}.

The number of required templates to cover a parameter space region $\Delta \lambda$ is then given by
\begin{equation}
    \mathcal{N}(\Delta \lambda) =
    \int_{\Delta \lambda} \mathrm{d} \lambda \prod_{k=1}^{D}\frac{1}{\delta \lambda[k]} \,.
    \label{eq:num_templates}
\end{equation}
where $D$ is the number of dimensions that can be resolved.
This expression ``counts'' the number of subdivisions of ``size'' $\delta \lambda$ required to cover
$\Delta \lambda$ for a non-constant $\delta \lambda$. Note that this prescription does not take into account
parameter-space correlations of the detection statistic (cf.~\cite{Wette:2014tca,Wette:2016raf,Prix:2006wm,
Messenger:2008ta, Messenger:2011rg, Leaci:2015bka}).

The relative importance of different parameter-space regions can be quantified by the ``local template density'' 
\begin{equation}
    \varrho(\lambda) =  \prod_{k=1}^{D}\frac{1}{\delta \lambda[k]}
\end{equation}
which can be used to guide template-placement algorithms. This quantity plays a similar role to
the \emph{volume element} derived from the determinant of the parameter-space metric~\cite{Prix:2006wm,
Prix:2007ks,Pletsch:2009uu, Messenger:2011rg,Wette:2013wza,Leaci:2015bka}.
In such setups, the parameter-space metric, which is related to Eq.~\eqref{eq:mismatch},
can be used to guarantee a parameter-space coverage with a maximum mismatch~\cite{Wette:2013wza,Wette:2016raf}. 
Here, on the other hand, $\varrho(\lambda)$ quantifies deviations in the frequency evolution. 
This is enough to distinguish templates, but requires the use of numerical methods to
understand the mismatch distribution and thus the sensitivity impact on a search, as we will discuss shortly.

As exemplified in Eq.~\eqref{eq:example}, $\varrho(\lambda)$ generally depends on parameter-space coordinates.
This makes problematic the application of standard template-placement methods based on lattice
quantization~\cite{Prix:2007ks, Wette:2013wza, Messenger:2011rg, Allen:2021yuy, Coogan:2022qxs, Agrell:2022jlo}.
We here propose a simple strategy. For usual all-sky searches (see Appendix~\ref{app:binary} for 
a worked example)~\cite{Krishnan:2004sv, Covas:2019jqa},  $\varrho(\lambda)$  is a monomial in $\lambda$ 
\begin{equation}
    \varrho(\lambda) = C \prod_{k=1}^{D} \lambda[k]^{p_{k}} \,,
    \label{eq:lambda_separable}
\end{equation}
where the $p_{k}$ exponents depend on the specific CW model and $C$ is an overall constant.
As a result, we can define a new set of coordinates $\unif(\lambda)$ so that $\varrho(\xi) = 1$:
\begin{equation}
    \mathrm{d}\lambda \,  \varrho(\lambda) = \mathrm{d}\unif \,.
    \label{eq:def_lambda}
\end{equation}
Effectively, this means sampling \emph{uniformly} on the $\unif$
space corresponds to sampling according to $\varrho(\lambda)$ in $\lambda$. 

By means of Eq.~\eqref{eq:lambda_separable}, we can split Eq.~\eqref{eq:def_lambda} into component-wise equations
\begin{equation} 
    \frac{\mathrm{d}\unif[k]}{\mathrm{d}\lambda[k]} = \sqrt[D]{C} \lambda[k]^{p_k} \,,
\end{equation}
which implies
\begin{equation}
    \xi[k] = \frac{\sqrt[D]{C}}{p_k + 1} \lambda[k]^{p_k + 1} \,.
    \label{eq:xi_of_lambda}
\end{equation}
This defines a bijection between the physical parametrization $\lambda$ and
the \emph{uniform-density} parametrization $\xi$ whose inversion is trivial. 

The generation of a template bank for a non-uniform density $\varrho(\lambda)$ such as that described in Eq.~\eqref{eq:lambda_separable}
goes then as follows: For a given parameter space region $\Delta \lambda$,
\begin{enumerate}
    \item Map the boundaries in $\Delta \lambda$ to the boundaries in $\xi$ using Eq.~\eqref{eq:xi_of_lambda}:
    \begin{align*}
        \xi_{\min} &= \xi(\lambda_{\min})\\
        \xi_{\max} &= \xi(\lambda_{\max})
    \end{align*}
    \item Generate a template bank in $\xi$ according to any prescription.
    \item Map the template bank in $\xi$ back to $\lambda$ using Eq.~\eqref{eq:xi_of_lambda}.
\end{enumerate}

\subsubsection{Random template bank}

$\xi$ coordinates allow for standard lattice template-bank construction methods~\cite{Prix:2007ks, Wette:2013wza}.
Here, on the other hand, we explore the use of \emph{random} template banks~\cite{Messenger:2008ta, Allen:2022lqr}.
These are easy to implement in our setup and play nicely with the batch-evaluation of templates: Instead of walking
a grid, simply draw  $n$ templates uniformly in $\xi$, map them to $\lambda$, and evaluate the
detection statistic using $\texttt{vmap}\, s$. The expected sensitivity loss of a random template bank compared to that
of an optimal quantize tends to decrease with the parameter-space dimensionality~\cite{Allen:2021yuy}. This makes them
well suited for our work, as we will be ultimately interested in the problem of unknown CW sources in binary systems
\mbox{($D \geq 6$)}.

We characterize random template banks in this setup by means of the \emph{oversampling} $o$, which corresponds to the
ratio of templates in the template bank over the number of templates in the covered region according to Eq.~\eqref{eq:num_templates}.

\begin{figure}
    \includegraphics[width=\columnwidth]{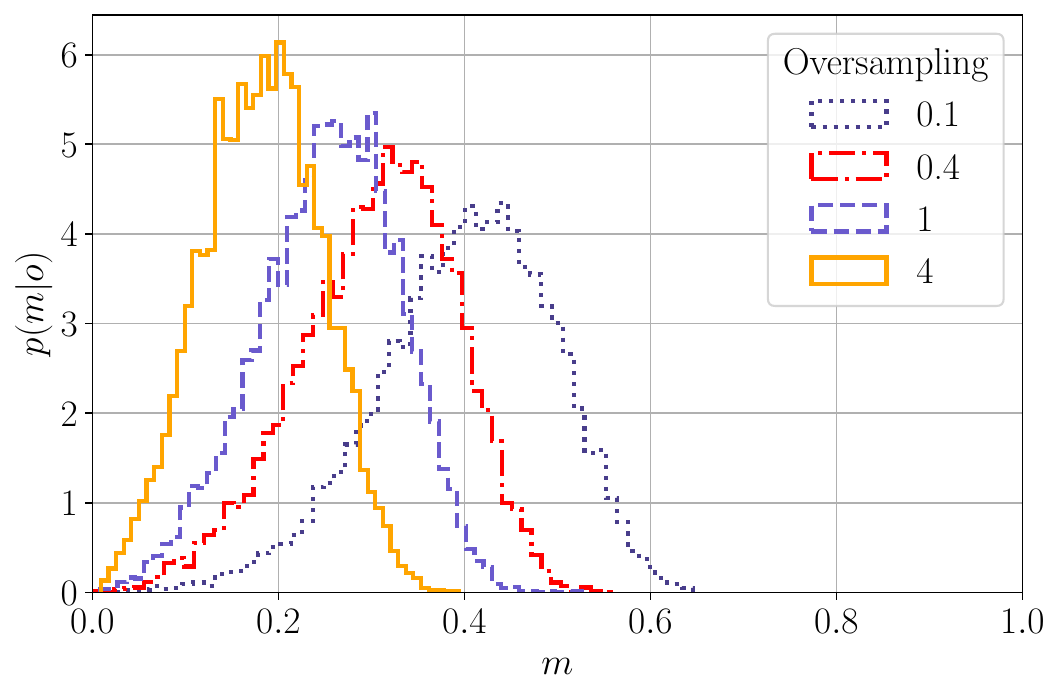}
    \caption{
        Mismatch distributions for the search setup described in Sec.~\ref{sec:real_data} using Gaussian noise.
        The results shown here correspond to a frequency of about $\SI{100}{\hertz}$; comparable results are obtained
        for other frequency bands.
    }
    \label{fig:mismatches}
\end{figure}

For a given parameter-space region $\Delta \lambda$, the mismatch distribution $p(m|o)$ can be sampled as follows:
\begin{enumerate}
    \item Generate a Gaussian-noise data stream containing a CW signal according to the search's priors.
    \item Uniformly sample $o \mathcal{N}(\Delta \lambda)$ templates in $\xi$ space and map them to $\lambda$ space.
    \item Evaluate $\texttt{vmap}\, s$ on the generated templates and compute $m$ according to Eq.~\eqref{eq:mismatch}.
\end{enumerate}  
We show the mismatch distribution corresponding to the search setup described in Sec.~\ref{sec:real_data} for different
oversampling values in Fig.~\ref{fig:mismatches}. Concretely, templates are uniformly distributed in a narrow frequency band 
of $\SI{0.125}{\hertz}$, across the sky, and across the binary parameter-space region described in Sec.~\ref{sec:real_data}. 
We use simulated Gaussian noise in a consistent manner with the dataset described in Sec.~\ref{sec:real_data}.
These results correspond to a frequency of about $\SI{100}{\hertz}$;
different frequency bands yield similar results, as expected from the definition of $\xi$. As we shall discuss in 
Sec.~\ref{sec:sensitivity_estimate}, these distributions allow us to connect computing-cost related quantities
(such as oversampling $o$) to sensitivity-related quantities (such as mismatch $m$).
The results seem to be compatible with Weibull distributions for average mismatch values away from $m=0$ and $m=1$.
This is compatible with the expected result from extreme value theory~\cite{Tenorio:2021wad}. For increasing values of $o$,
the mismatch reduction slows down with respect to $o$, as discussed in~\cite{Messenger:2008ta}.

\subsection{Ahead-of-time clustering\label{subsec:aot_clustering}}

As previously discussed, post-processing strategies~\cite{Singh:2017kss,Morawski_2020,Tenorio:2020cqm,Beheshtipour:2020zhb,
Beheshtipour:2020nko,Steltner:2022aze} work by grouping together nearby candidates in the parameter space,
usually by means of a clustering algorithm. This has the effect of reducing the detectability threshold at
a given follow-up cost~\cite{Mirasola:2024lcq}. Ultimately, however, the operative result of these algorithms 
is to select \emph{uncorrelated local maxima} across the parameter space. One could propose a simpler 
post-processing approach  by splitting the parameter-space into disjoint regions so that the search step consists
in retrieving the loudest candidate within each region.

Meaningfully dividing the parameter space is in general difficult due to the dependency of the parameter-space 
metric with  the parameter-space position itself. We here propose a simpler approach using the $\xi$ coordinates
to divide region $\Delta \lambda$ into disjoint hyperboxes with a constant number of templates. 
To do so, note that, by construction,
\begin{equation}
    \mathcal{N}(\Delta \lambda) = \prod_{k=1}^{D} \Delta \xi[k] \,,
\end{equation}
i.e. the number of templates in a region $\Delta \lambda$ is the \emph{coordinate volume} enclosed in $\xi$.
Boxes with a fixed number of templates, say $n_{\mathrm{b}}$, can be created by partitioning $\Delta \xi$ into regular 
hyperboxes with a coordinate volume of $n_{\mathrm{b}}$ templates. This approach is significantly simpler than a 
clustering, and could behave in a similar manner if hyperboxes are properly selected.

Statistically speaking, different boxes will have a different set of weights $w_{X \alpha}(\lambda)$ as they will contain
different templates $\lambda$. Since we work in the \mbox{$\Nsft \gg 1$} limit, in which $s$ follows a Gaussian distribution,
it is enough to standardize the detection statistic to a zero-mean unit-variance Gaussian distribution to make comparisons
across different boxes
\begin{equation}
    \crit(\lambda;b) = \frac{s(\lambda) - \mu_{\mathrm{G}}(b)}{\sigma_{\mathrm{G}}(b)} \,.
    \label{eq:critical_ratio}
\end{equation}

At this point, is up to the search setup to decide how many boxes to select. Note that \emph{any} selection scheme is allowed,
including deliberately rejecting boxes containing instrumental artifacts.
The effect of rejecting a certain number of boxes
can be negligible, if the number of boxes is high enough, as sensitivity estimates ultimately operate
at 90\% to 95\% detection probability~\cite{Tenorio:2021wmz}. To decide the number of partitions,
we ensure a minimum of 10,000 boxes in any narrow search frequency band (typically covering 
about $\SI{0.1}{\hertz}$) to allow for 
the rejection of highly contaminated regions, and cap the box size in such a way that
the number of templates results in an efficient evaluation given the available GPUs at hand.

This ahead-of-time partition provides a unified picture of the post-processing algorithms used in short-coherent blind CW searches,
as well as a simple and computationally efficient proposal of post-processing, as the memory, computing cost, and tuning parameters
have been significantly reduced compare to previous approaches.

\section{Sensitivity of a short-coherence search\label{sec:sensitivity_estimate}}

The sensitivity of a blind CW search is typically expressed in terms of the minimal amplitude $h_0$ so that 
a fraction $\pdet$ of an isotropically-oriented all-sky population of sources is detected~\cite{Tenorio:2021wmz,Wette:2023dom}. 
Typically, instead of $h_0$ the results are stated in terms of the \emph{sensitivity depth}~\cite{Behnke:2014tma, Dreissigacker:2018afk}
\begin{equation}
    \depth = \frac{\sqrt{\Sn}}{h_0} \,,
\end{equation}
which ends up a function of the search setup (the available dataset, the template-bank setup, 
and the detection statistic on use) and the signal population under consideration. 

In an actual search, the estimation of $\depth$ is done numerically though the use of software-simulated signals (injections),
which has an associated computing cost.  This motivated the development of semi-analytic sensitivity estimate
methods~\cite{Wette:2011eu, Dreissigacker:2018afk, Mirasola:2024lcq}, which as implemented in~\cite{octapps, cows3}
have a negligible cost and can be used as a first step to guide in-depth injection campaigns
These methods, however, are developed specifically for the $\mathcal{F}$-statistic (or the associated number 
count~\cite{Dreissigacker:2018afk}); closed-form estimates are also available for power-based
number-count~\cite{Krishnan:2004sv,2014PhRvD..90d2002A,KAGRA:2022osp, KAGRA:2022dwb}. These methods, however,
do not account for the effect of post-processing steps and weighted statistics.

Sensitivity estimates are ultimately rooted in the principles discussed in~\cite{Searle:2008jv}: 
Given a search setup and signal distribution
configuration $c$, the detection probability corresponds to the fraction of datasets $x$ in the data space that 
satisfy a certain detection rule $R(x;c)$: 
\begin{equation}
    \pdet(c) = \int \mathrm{d}x \, R(x; c) \, p(x|c) \,.
    \label{eq:pdet_abstract}
\end{equation}
$p(x|c)$ represents a data generating process according to the specifications in $c$, such as a CW signal in Gaussian noise.
If we fix a search setup and signal distribution, the dependency in $x$ of Eq.~\eqref{eq:pdet_abstract} can be recast
in terms of the detection statistic $z$ and the parameter-space region at hand as 
\begin{equation}
    \pdet(\depth, o, \{\thresh\}) = \int\mathrm{d}\lambda  \int \mathrm{d}\crit \, 
    R(\crit; \thresh(\lambda))  \, p(\crit | \depth, o ,\lambda) \, p(\lambda)\,,
    \label{eq:simple_eq}
\end{equation}
where now the configuration $c$ is given by the sensitivity depth of the population $\depth$, the oversampling of the template bank $o$
(see Sec.~\ref{subsec:uniform_tb}), and   the set of detectability thresholds $\{ \thresh \}$, which may depend on the 
parameter-space location as per Sec.~\ref{subsec:aot_clustering}. We discuss the definition of $p(\crit | \depth, o ,\lambda)$ in
Sec.~\ref{subsec:simulate}.

With this, estimating the sensitivity of a search, represented by the sensitivity depth $\depth^{p}$ at which 
a fraction $p$ of the population is detected becomes solving the one-dimensional equation
\begin{equation}
    \pdet(\depth^{p}, o, \{ \thresh\}) = p \,.
     \label{eq:solve_this}
\end{equation}
In practice, $\pdet$ is computed through Monte Carlo integration
\begin{equation}
    \begin{aligned}
    \pdet(\depth, o, \{\thresh\}) &\approx \frac{1}{N_\mathrm{I}}\sum_{i=1}^{N_\mathrm{I}} 
    R(\crit^{(i)}; \thresh(\lambda^{(i)})) \\
    \crit^{(i)} &\sim p(\crit | \depth, o ,\lambda^{(i)})\\
    \lambda^{(i)} &\sim p(\lambda)
\end{aligned}\,.
\label{eq:sampling}
\end{equation}
This calculations amount to computing what fraction of $N_{\mathrm{I}}$ software-simulated signals
is detected by the search pipeline. The process of simulating and running the search pipeline (i.e. computing the detection statistic)
corresponds in Eq.~\eqref{eq:sampling} to sampling $z^{(i)}$, whereas deciding whether said signal is detected
corresponds to evaluating $R(z^{(i)}; \thresh(\lambda^{(i)}))$. 

The basis for the semi-analytical sensitivity estimation algorithm proposed
in~\cite{Wette:2011eu, Dreissigacker:2018afk} is to derive an efficient formulation of  $p(\crit | \depth, o ,\lambda)$
so that it can be efficiently sampled. We generalize such an approach to short-coherence detection statistics in
Sec.~\ref{subsec:simulate} and the corresponding implementation of the detection rule $R(\crit;\thresh(\lambda))$
in Sec.~\ref{subsec:detection_rule}. The method is summarized in Sec.~\ref{subsec:pdet_summary}.

\subsection{Simulating detection statistics\label{subsec:simulate}}

We now discuss a simple procedure to draw samples from $p(\crit | \depth, o ,\lambda)$.
Note that  given parameter-space partition around a template $\lambda$ with known 
$\mu_{\mathrm{G}}$ and $\sigma_{\mathrm{G}}$,  sampling  $p(\crit | \depth, o ,\lambda)$ is equivalent to sampling
$p(s | \depth, o ,\lambda)$ and then computing $z = (s - \mu_{\mathrm{G}}) / \sigma_{\mathrm{G}}$. As a result 
we work in terms of $s(\lambda)$ to directly apply the results derived in Sec.~\ref{sec:statistics}.

We start by introducing explicit dependencies on the mismatch and the orientation angles
in $p(s|\depth, o, \lambda)$.
The reason is that $s$ can be sampled using Eq.~\eqref{eq:s_gauss} as long as we fix 
a specific parameter-space region and we
know the depth $\depth$, mismatch $m$ and pair orientation angles $\hA$.
To do so, marginalize over the mismatch 
$m$ and orientation angles $\hA$:
\begin{equation}
    p(s|\depth, o, \lambda) = \int \mathrm{d}\hA \int \mathrm{d}m \, p(s|\depth, m, \lambda, \hA) \, p(\hA) \, p(m|o) \,,
\end{equation}
$p(m|o)$ can be obtained numerically at a negligible cost as discussed  in Sec.~\ref{subsec:uniform_tb}. 
The distribution $p(s|\depth, m, \lambda, \hA)$ corresponds to Eq.~\eqref{eq:s_gauss}.
We now proceed in a similar manner to~\cite{Wette:2011eu, Dreissigacker:2018afk} and recognize that
Eq.~\eqref{eq:s_gauss} depends on $\hA$, $\lambda$, and the observing run configuration through 
\emph{only 4 parameters}  \mbox{$q = \{\mu_{\mathrm{G}}, \sigma_{\mathrm{G}}, \hat{\rho}^2_1, \hat{\rho}^2_2\}$},
As a result, we can apply a similar argument once more to obtain
\begin{equation}
    p(s|\depth, o, \lambda) = \int \mathrm{d}m \int \mathrm{d}q \, p(s|\depth, m, q) \, p(q|\lambda) \, p(m|o) \,,
\end{equation}
where the dependency on $\lambda$ is contained within the weights $w_{X\alpha}$ inside $q$.

The four quantities in $q$ play an analogous role to the ``response function'' and similar quantities governing the 
distribution of the $\mathcal{F}$-statistic~\cite{Wette:2011eu, Dreissigacker:2018afk, Mirasola:2024lcq}.
These four quantities are due to 1) the Gaussian distribution of $s$ in the limit \mbox{$\Nsft \gg 1$},
which requires only \emph{two} parameters to be characterized, and 2) the use of weights, 
which modify the distribution parameters under \emph{both} the signal and noise hypotheses. 
Similar approaches are applicable for the treatment of other weighted statistics
such as~\cite{Covas:2022mqo,Prix:2024xpl}.

Specifically, $q$ is a set of deterministic function of the orientation parameters $\hA$ and 
the sky position $\hat{n}$.  Given a specific signal population, which fixes $p(\hA, \hat{n})$, we can numerically generate the distribution of $q$ \emph{once for a given observing run} as follows: First, we draw a set of sky-position
and amplitude parameters from the population priors
\begin{equation}
    \hA^{(i)}, \hat{n}^{(i)} \sim p(\hA, \hat{n}) \,.
\end{equation}
These are combined with the SFT timestamps and antenna-pattern functions to compute the per-SFT data-dependent
quantities
\begin{equation}
\begin{aligned}
    \hat{\rho}^{2 \, (i)}_{X\alpha} &= \hat{\rho}^2_{X\alpha}(\hA^{(i)}, \hat{n}^{(i)}) \\
    w_{X\alpha}^{(i)} &= w_{X\alpha}(\hat{n}^{(i)})
\end{aligned}\,,
\end{equation}
which completely determine the four quantities in $q$:
\begin{equation}    
\begin{aligned}
        \mu_{\mathrm{G}}^{(i)} &= (1.012) \times 2 \sum_{X, \alpha} w^{(i)}_{X\alpha} \\
        \sigma_{\mathrm{G}}^{2\, (i)} &= 4 \sum_{X, \alpha} (w^{(i)}_{X \alpha})^2\\
        \hat{\rho}^{2\, (i)}_{1} &= \sum_{X,\alpha}  w^{(i)}_{X \alpha} \hat{\rho}^{2 \, (i)}_{X\alpha}\\
        \hat{\rho}^{2\, (i)}_{2} &= \sum_{X,\alpha}  (w^{(i)}_{X \alpha})^2 \hat{\rho}^{2 \, (i)}_{X\alpha}\\
        \end{aligned} \,.
\end{equation}
where the extra 1.012 factor in $\mu_{\mathrm{G}}^{(i)}$ corresponds to the PSD-estimation bias, as discussed
in Sec.~\ref{sec:statistics}.

\subsection{Detection rules\label{subsec:detection_rule}}

The second required ingredient is the detection rule, which is a function which decides whether
a data realization contains a signal or not. In a CW search, this function encapsulates the effect of post-processing steps
and other significance vetoes imposed in the resulting parameters. As a result, it is typically difficult to model 
and most sensitivity estimation algorithms approximate it as a step function depending on the detectability threshold. 

The specific post-processing proposed in Sec.~\ref{subsec:aot_clustering}, however, allows for a simple modeling: 
The threshold against which a candidate is compared is a function of the parameter-space box in which said template falls. 
This corresponds to defining a function $\thresh(\lambda)$ which returns
the corresponding detectability threshold, be it in terms of $s$ or $z$. As a result, a general detection rule accounting for the
effects our post-processing can be readily formulated as
\begin{equation}
    R(z; \thresh(\lambda)) = \left\{ \begin{matrix} 1 & \mathrm{if} \,  z > \thresh(\lambda)\\ 0 & \mathrm{otherwise} \end{matrix}\right\} \,.
\end{equation}

\subsection{Summary\label{subsec:pdet_summary}}

Equation~\eqref{eq:solve_this} can be easily solved by numerically computing $\pdet$ using a Monte Carlo integral as follows:
\begin{equation}
\begin{aligned}
    \pdet(\depth, o, \{t\}) &\approx \frac{1}{N_\mathrm{I}}\sum_{i=1}^{N_\mathrm{I}} 
    \left\{ \begin{matrix} 1 & \mathrm{if} \,  z^{(i) } > \thresh(\lambda^{(i)})\\ 0 & \mathrm{otherwise} \end{matrix}\right\}\, \\ç
    z^{(i)} &= \left[ s^{(i)} - \mu_{\mathrm{G}}(\lambda^{(i)}) \right] / \sigma_{\mathrm{G}}(\lambda^{(i)}) \\
    s^{(i)} &\sim  p(s | \depth, m^{(i)},  q^{(i)}) \\
    q^{(i)} &\sim p(q | \lambda^{(i)}) \\
    \lambda^{(i)} &\sim p(\lambda)\\
     m^{(i)} &\sim p(m|o)\\
\end{aligned}\,.
\end{equation}
The distributions associated to $q$ and $m$ are independent of $\depth$ and well-determined for a given search setup and sky-distribution.
As a result, the evaluation of $\pdet(\depth, o, \{\thresh\})$ can be efficiently computed using any array-capable computing software.
We release an implementation of $\pdet$ under the \texttt{cows3} package~\cite{Mirasola:2024lcq, cows3}.

Let us summarize the main result of this section: Sampling $p(z|\mathcal{D}, o, \lambda)$ is statistically
equivalent to injecting a CW signal in Gaussian noise at depth $\depth$ with orientation and sky 
location parameters drawn from $p(\hA, \hat{n})$ and retrieving the maximum detection statistic $z$ 
from a random template bank with oversampling $o$. The computing gains come from the fact that, once 
the distribution of $q$ is computed, sampling $p(z|\mathcal{D}, o, \lambda)$ is several orders of magnitude
faster than generating software-simulated signals.

Note that, for this procedure to be applicable, $p(z|\depth, o, \lambda)$ only needs to be \emph{sampled},
in a similar manner to  the ``synthetization'' approach~\cite{Prix:2007ks,Prix:2011qv,Prix:2024xpl}. As a result,
this approach is applicable to a broad array of detection 
statistics~\cite{Prix:2007ks,Prix:2011qv,Keitel:2013wga,Keitel:2015ova,Covas:2022xyd,Prix:2024xpl}.

\section{Applications to real data\label{sec:real_data}}

We test the effectiveness of these developments on data from the first half of the LVK's third observing 
run~\cite{KAGRA:2023pio}. We deploy an all-sky search for unknown CW sources in binary systems in a selection
of frequency bands, a setup similar to~\cite{LIGOScientific:2020qhb}: We select 8 representative
\SI{0.125}{\hertz} frequency bands ([110.500,
    132.625,
    168.000,
    187.000,
    207.125,
    228.625,
    252.625,
    260.625] Hz)
where we deploy an all-sky search covering orbital periods of 7 to 15 days and 
projected semi-major axes of 5 to 15 light-seconds. We use SFTs with a duration of $\Tsft = \SI{1024}{\second}$
with a 50\%  overlap and windowed by a Tukey widow with a 0.5 tapering parameter~\cite{Astone:2000jz}.
The results will be presented in terms of population-based sensitivity estimates, which will be compared to
the sensitivity estimates produced by the method in Sec.~\ref{sec:sensitivity_estimate}.

\subsection{Search setup}

The aim of this study is to showcase the applicability of the simple approaches presented throughout this work.
To do so, we deploy an all-sky search in real data and compute the corresponding sensitivity estimates using different
search setups. We assume that all the selected outliers would be followed-up and discarded by a more sensitive method
such as~\cite{Ashton:2018qth, Tenorio:2021njf, Covas:2024pam, Mirasola:2024lcq}. The results will be 
informative as they will be equivalent to those obtained from an actual search if no signal is detected.

For each frequency, we deploy a template bank with using an oversampling of $o=3.44$.
This corresponds to about 1 million GPU computing hours to cover the full frequency band from \SI{100}{\hertz} to \SI{300}{\hertz}.
We divide the parameter space in each band into hyperboxes containing about $10^{8}$ templates, ensuring always a minimum of
10,000 boxes per band. To account for non-Gaussianities in the data, we compute detection statistics
templates as discussed in Sec.~\ref{subsec:particularities} and retrieve the loudest template per box. 

Upon completion of the main search stage we are left with a list of templates and their corresponding detection statistic
$\mathcal{T} = \{ (\lambda_{b}, s_{b}), b = 1, \dots, N_{\mathrm{boxes}} \}$. In line with standard
practice~\cite{Tenorio:2021njf, Covas:2024pam, Mirasola:2024lcq}, a certain number of candidates $N_{\mathrm{cand}}$ is selected
to follow-up. This procedure sets the corresponding detectability thresholds for the sensitivity estimate procedure.
We test here two different post-processing approaches which intend to be comparable to those used in current post-processing
algorithms~\cite{Tenorio:2021wmz}.

The first approach is to select the top $N_{\mathrm{cand}}$ candidates in $\mathcal{T}$.
This implies the detectability of a signal
depends on the hyperbox where it falls: If the signal falls within a hyperbox containing a selected candidate,
said signal is detected if it has a detection statistic above that of the candidate; otherwise, 
a signal is detected if it has a greater detection statistic than the last selected candidate.

A second approach makes use of the sensitivity procedure in Sec.~\ref{sec:sensitivity_estimate}. 
Having access to $\mathcal{T}$ implies
we can numerically evaluate $\depth^{95\%}$ as a function of $N_{\mathrm{cand}}$ very efficiently. As a result, 
we can compute the outlier-selection strategy that  \emph{maximizes} $\depth^{95\%}$ for a given $\mathcal{T}$.
To do so, we sort the candidates in $\mathcal{T}$ in descending order according to their detection statistic.
We then allow $N_{\mathrm{reject}}$ hyperboxes to be \emph{rejected} and $N_{\mathrm{cand}}$ hyperboxes to be selected.
This allows for highly-disturbed regions containing noise artifacts to be ignored so that computing resources
are better employed, in a similar manner to what was argued in~\cite{Mirasola:2024lcq}.
The resulting detection rule is similar to that of the previous approach, except that signals in an ignored hyperbox are not detected. Note that this may give rise to diminishing returns, as rejecting too many boxes may end up hindering
the search's sensitivity.

We make use of this last approach, using the sensitivity estimate strategy developed in
Sec.~\ref{sec:sensitivity_estimate} to optimize the number of rejected and selected boxes.
This procedure is computationally very efficient, as it only involves evaluating a handful of
Monte Carlo integrals, and amounts to less than 1 minute on a laptop for a given frequency band.

The actual sensitivity of these different setups is then evaluated by means of an injection campaign.
We inject 500 simulated signals at different, discrete depth values. 
These injections are isotropically oriented and uniformly distributed in the sky and the binary
parameters~\cite{LIGOScientific:2020qhb}. From these, we compute the detected fraction of the population at a given depth.

\subsection{Particularities of real-data searches\label{subsec:particularities}}

Noise in a real-data search is generally non-Gaussian and contains different kinds of instrumental artifacts that degrade the
sensitivity of a search~\cite{2018PhRvD..97h2002C}. This generally causes the results derived in Sec.~\ref{sec:sensitivity_estimate}
to be unaplicable, as the underlying distributions of $s$ are practically unknown. One approach to overcome the effect of
instrumental artifacts in CW searches is to design better suited detection statistics, such as~\cite{Keitel:2013wga,Keitel:2015ova}.
Here, we follow a simpler approach proposed in~\cite{Covas} using multiple detection statistics.

Non-Gaussianities are capable of accumulating a large amount of normalized power in a relatively ``short'' period of time~\footnote{Note that non-Gaussianities do not need to be short themselves; 
rather, it is enough that the frequency-evolution track of a signal overlaps
such non-Gaussianity for a short period of time.}.
As a result, the loudest template in a box can either be caused by a genuine CW signal or
by a non-Gaussianity.  The key property to tell these two phenomena apart relies on
quantifying their \emph{persistence}. This was initially discussed in~\cite{Krishnan:2004sv}
which proposed the use of the \emph{number count} statistic.  The weighted number count of a 
signal is compute using the binarized spectrogram according to some threshold, and provides an 
adequate notion of persistence.


We proceed in a similar manner to the approach proposed in~\cite{Covas}: We retrieve the candidate with the \emph{loudest}
normalized power with the extra restriction of scoring \emph{at least} a pre-established weighted number count value.
We choose said value to be three standard deviations above the average number count in Gaussian noise, based on our numerical results.
Statistics of the weighted number count are derived in~\cite{wnc}.

\subsection{Results}

\begin{figure}
    \includegraphics[width=\columnwidth]{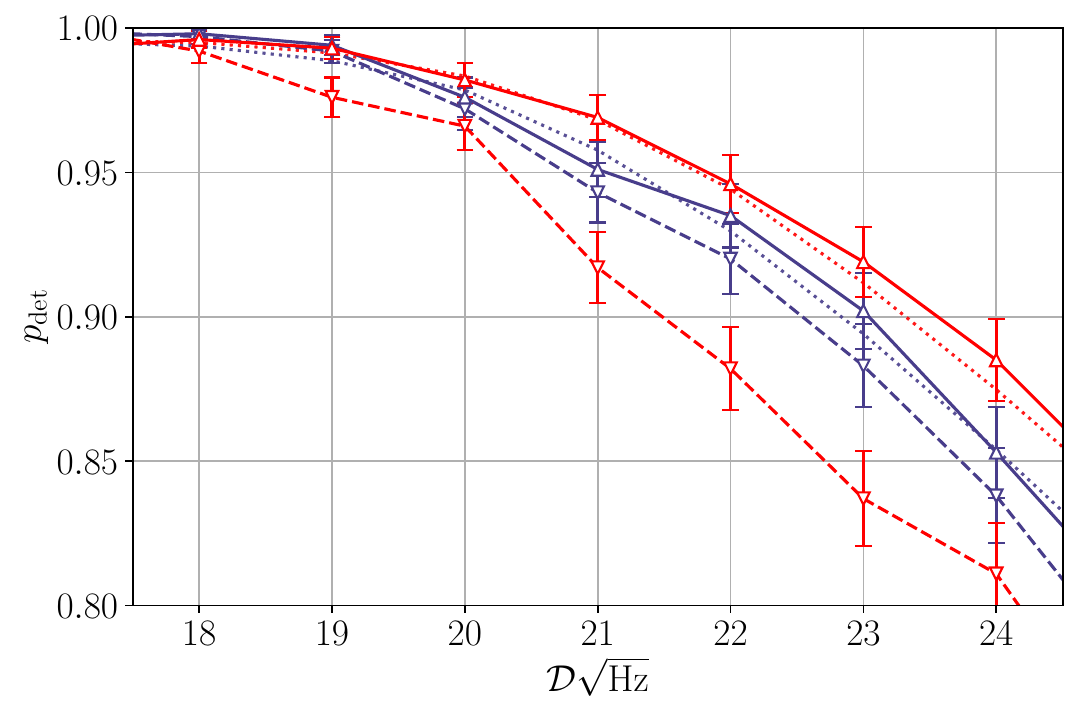}
    \caption{
        Detection probabilities at different sensitivity depths for the signal population discussed in the text
        at \SI{110.5}{\hertz}.
        Each line corresponds to a different detection statistic and post-processing configuration.
        The blue lines correspond to retrieving the template with maximum normalized power without a minimal weighted number count;
        the red lines impose a minimal weighted number count as discussed in the text.
        The dashed lines correspond to $N_{\mathrm{cand}} = 1$. 
        The solid lines correspond to the optimal choice of $N_{\mathrm{reject}}$ and $N_{\mathrm{cand}}$.
        The dotted lines show the estimated sensitivity for said optimal choice using the method from Sec.~\ref{sec:sensitivity_estimate}
    }
    \label{fig:1105_comparison}
\end{figure}

We start by comparing these different approaches for the \SI{110.5}{\hertz} frequency band in Fig.~\ref{fig:1105_comparison}.
While for a low number of selected candidates $N_{\mathrm{cand}} = 1$ both approaches behave similarly, the use of a number-count
filter significantly improves the detection probability towards higher depths (i.e. lower signal amplitudes) 
for the optimal candidate selection setup compared to only using normalized power. Furthermore, if a minimal number count is imposed,
the measured detection probability is in agreement with the estimate provided by the method in Sec.~\ref{sec:sensitivity_estimate}.
This supports the idea that requiring a minimal persistency tames the effect of non-Gaussianities of the data. 

\begin{figure}
    \includegraphics[width=\columnwidth]{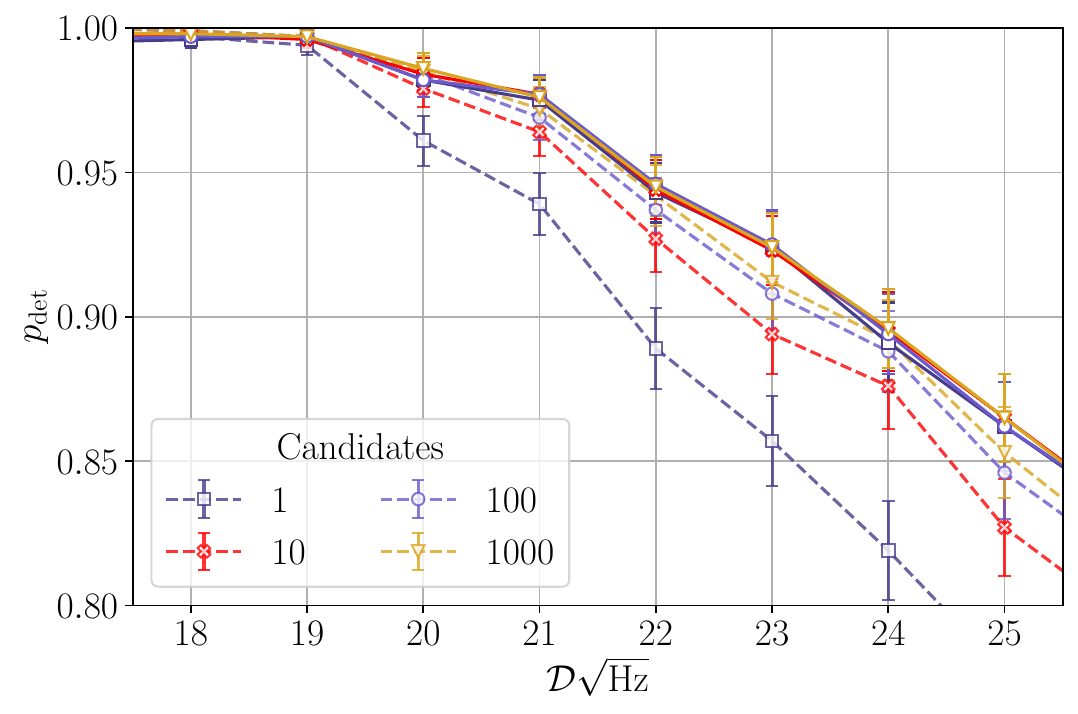}
    \caption{
        Comparison of detection probabilities for different number of selected candidates $N_{\mathrm{cand}}$
        imposing a minimal number count at \SI{110.5}{\hertz}. 
        Dashed lines correspond to $N_{\mathrm{reject}} = 0$, while solid lines 
        use the optimal number of rejected boxes.
    }
    \label{fig:1105_how_many}
\end{figure}

After identifying the use of number count as a favorable choice, we show in Fig.~\ref{fig:1105_how_many} the effect of rejecting
a certain number of boxes. Specifically, we compare the detection probabilities for $N_{\mathrm{cand}} = 1, 10, 100, 1000$ without
rejecting boxes and using the optimal number of rejected boxes for each number of candidates. We observe that, for this specific case, 
the sensitivity depends weakly on $N_{\mathrm{cand}}$ once the optimal number of boxes has been rejected. 
This is consistent with the results of~\cite{Mirasola:2024lcq}:  Loud candidates are caused by a mixture of non-Gaussianities 
in the data and a Gaussian background.
Non-Gaussianities tend to cause highly-correlated elevated normalized power structures that are well contained within a certain 
number of hyperboxes; extreme events caused by a Gaussian background, on the other hand, are scattered across the parameter
space and scale weakly with the number of selected candidates~\cite{Tenorio:2021wad}. As a result, the majority of the 
sensitivity gain is achieved upon rejecting all hyperboxes containing non Gaussianities.

\begin{figure}
    \includegraphics[width=\columnwidth]{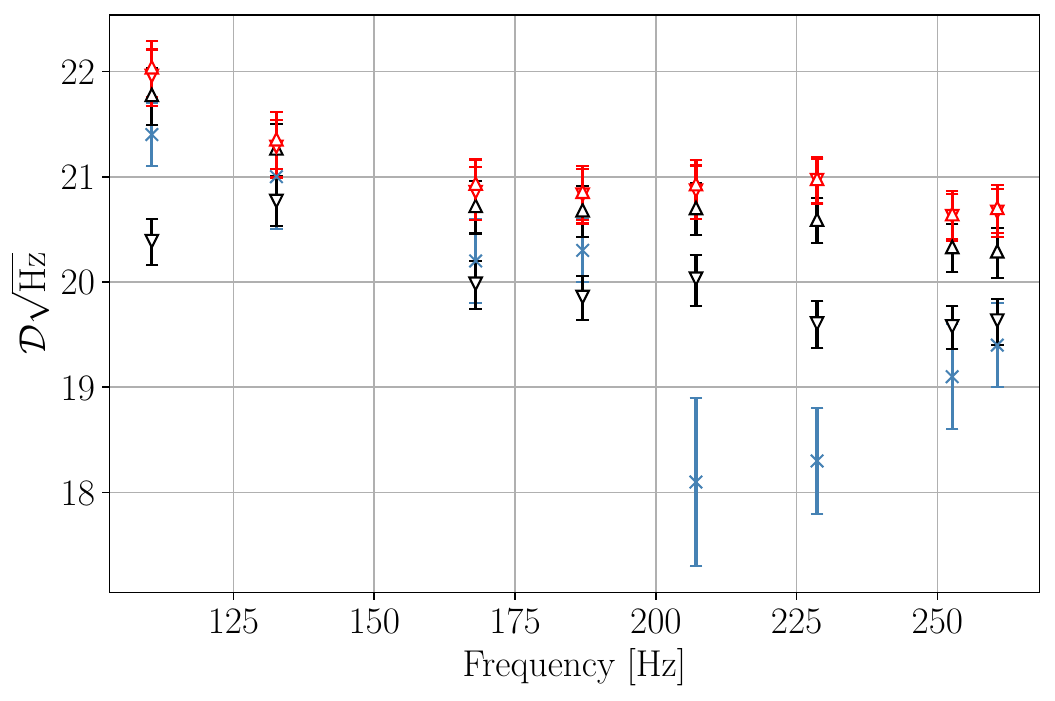}
    \caption{
    Sensitivity depth corresponding to a 95\% detection probability for eight representative frequency bands.
    All results are computed imposing a minimum number count as discussed in the main text.
    Downward-pointing triangles correspond to \mbox{$N_{\mathrm{cand}} = 1$}, while upward-pointing triangles
    correspond to \mbox{$N_{\mathrm{cand}} = 1000$}. Black markers correspond to \mbox{$N_{\mathrm{reject}} = 0$}, 
    while red markers correspond to choosing the optimal \mbox{$N_{\mathrm{reject}}$} for the chosen 
    number of candidates. The blue crosses correspond to the results quoted in~\cite{LIGOScientific:2020qhb} and
    are only shown for completeness. 
    }
    \label{fig:depth_vs_freq}
\end{figure}

We produce similar results for seven other frequency bands and compute the sensitivity depth at which 95\% of the signals are detected.
To do so, we fit a sigmoid to the detection probability and compute the corresponding uncertainty through the covariance matrix of the
fit, as discussed in~\cite{LIGOScientific:2020qhb}. The results are shown in Fig.~\ref{fig:depth_vs_freq}. We observe compatible results
with those displayed by the \SI{110.5}{\hertz} band: Upon selecting an optimal number of rejected boxes, the resulting sensitivity
is weakly dependent on the number of candidates to follow-up. 
The specific gain depends on the properties of the background noise,
which tends to behave differently depending on the frequency band.

The sensitivity depths here obtained are broadly consistent with those reported in~\cite{LIGOScientific:2020qhb}.
This is expected, as both setups use the same detection statistics and $\Tsft$. In light of this result,
we can conclude that the developments presented in this work are a feasible proposal to conduct broad
parameter-space searches in future observing runs.

\section{Conclusion\label{sec:conclusion}}

The search for CW signals from unknown sources is a formidable computational challenge
due to the involved parameter-space breadths. To tackle such a problem, multiple detection
pipelines benefiting from different parameter-space properties have been developed in order
to efficiently cover large portions of the parameter space~\cite{Jaranowski:1998qm, Astone:2000jz, Dergachev:2010tm, Dergachev:2019wqa,
Krishnan:2004sv, 2014PhRvD..90d2002A, Miller:2018rbg, Oliver:2019ksl, Covas:2019jqa, Covas:2022mqo, powerflux, powerflux2}.

In this work, we have studied the performance of a general GPU-accelerated search, \fasttracks{},
which we release as an open-source Python package~\cite{fasttracks}.
We find that the simple brute-force implementation of short-coherence detection statistics using
GPU-capable array language~\cite{jax, cupy_learningsys2017, 2019arXiv191201703P} provides a computational
efficiency comparable to that of GPU-accelerated pipelines for similar detection statistics,
such as~\cite{Covas:2019jqa, Rosa:2021ptb}, at a lower code complexity. This promising result will be
key to develop computationally-efficient methods to search for more realistic NS 
physics~\cite{Ashton:2017wui, Mukherjee:2017qme, Sieniawska:2022bcn}, as well as other, more exotic
CW sources~\cite{Isi:2018pzk,Sun:2019mqb,Palomba:2019vxe,
Zhu:2020tht,Jones:2023fzz,Miller:2022wxu,Miller:2022wxu,Miller:2023kkd,
KAGRA:2024ipf, Bhattacharya:2024pmp,Guo:2022sdd,Miller:2024khl,Miller:2024fpo}.
Similar strategies were used as part of the winning solutions
in a recent Kaggle data analysis competition for CW signals~\cite{kaggle}.

Furthermore, we have explored the formulation of template-bank setup and post-processing algorithms. 
Notably, we have proposed an alternative to clustering algorithms based on preemptively splitting the parameter space
into disjoint regions. This has the advantage of being simpler to implement and free from configuration choices such
as parameter-spaces distance of clustering-specific parameters. Moreover, this procedure is easy to model from a statistical
point of view. The feasibility of this approach has been successfully tested on a real-data application using data from the 
LVK's third observing run. These results suggest alternative post-processing algorithms, 
such as~\cite{Beheshtipour:2020nko, Beheshtipour:2020zhb}, may improve upon current clustering approaches in terms of
computing cost and computational complexity.

Finally, we have generalized the sensitivity estimate procedure presented in~\cite{Wette:2011eu, Dreissigacker:2018afk, octapps}
to short-coherence detection statistics and to account for post-processing strategies such as the ones here presented.
This will allow for a significant computing-cost reduction of all-sky sensitivity estimates and a better understanding of
the optimal set up of real-data searches. We release this algorithm as part of the \texttt{cows3}~\cite{Mirasola:2024lcq, cows3}
open-source package.

The success of these simple ideas in the search for CW signals will become increasingly relevant for the search
of long-duration GW signals such as those to come with next generation detectors~\cite{Maggiore:2019uih,Reitze:2019iox,
2017arXiv170200786A,Hu:2017mde,Li:2024rnk}, as shown by one of the authors in~\cite{Tenorio:2025yca}.

\section*{Acknowledgements}

We thank Rafel Jaume, David Keitel, Andrew Miller,
and the CW working group of the LIGO-Virgo-KAGRA Collaboration for comments on the manuscript.
We also thank Fréderic Bastien, Michel Herrera Sanchez, Mahmoud Solimand,
and Yu-Hang Tang for their support during the CINECA Open Hackathon 2023.
This work was supported by 
the Universitat de les Illes Balears (UIB);
the Spanish Agencia Estatal de Investigación grants 
PID2022-138626NB-I00, RED2022-134204-E, RED2022-134411-T, 
funded by MICIU/AEI/10.13039/501100011033 and the ERDF/EU;
and the Comunitat Autònoma de les Illes Balears through the Conselleria d'Educació i Universitats
with funds from the European Union - NextGenerationEU/PRTR-C17.I1 (SINCO2022/6719) and from 
the European Union - European Regional Development Fund (ERDF) (SINCO2022/18146).
RT is supported by 
ERC Starting Grant No.~945155--GWmining, 
Cariplo Foundation Grant No.~2021-0555, 
MUR PRIN Grant No.~2022-Z9X4XS, 
MUR Grant ``Progetto Dipartimenti di Eccellenza 2023-2027'' (BiCoQ),
and the ICSC National Research Centre funded by NextGenerationEU.
JRM is supported by the Spanish Ministerio de Ciencia, Innovación y Universidades (ref.~FPU 22/01187).
The authors are grateful for computing resources at
Artemisa, funded by the European Union ERDF and Comunitat Valenciana as well as the 
technical support provided by the Instituto de Fisica Corpuscular, IFIC (CSIC-UV);
MareNostrum5, as well as the technical support provided by Barcelona Supercomputing Center 
(RES-FI-2024-3-0013,RES-FI-2025-1-0022);
the Digital Research Alliance of Canada (alliancecan.ca);
and the LIGO Laboratory, which is supported by National Science Foundation Grants PHY-0757058 and PHY-0823459.
This material is based upon work supported by NSF's LIGO Laboratory which is a major facility fully
funded by the National Science Foundation
This research has made use of data or software obtained from the 
Gravitational Wave Open Science Center (gwosc.org),
a service of the LIGO Scientific Collaboration, the Virgo Collaboration, and KAGRA.
This material is based upon work supported by NSF's LIGO Laboratory 
which is a major facility fully funded by the
National Science Foundation,
as well as the Science and Technology Facilities Council (STFC) of the United Kingdom,
the Max-Planck-Society (MPS), and the State of Niedersachsen/Germany for support of the 
construction of Advanced LIGO and construction and operation of the GEO600 detector.
Additional support for Advanced LIGO was provided by the Australian Research Council.
Virgo is funded, through the European Gravitational Observatory (EGO),
by the French Centre National de Recherche Scientifique (CNRS),
the Italian Istituto Nazionale di Fisica Nucleare (INFN) and the Dutch Nikhef,
with contributions by institutions from Belgium, Germany, Greece, Hungary,
Ireland, Japan, Monaco, Poland, Portugal, Spain.
KAGRA is supported by Ministry of Education, Culture, Sports,
Science and Technology (MEXT),
Japan Society for the Promotion of Science (JSPS) in Japan;
National Research Foundation (NRF) and Ministry of Science and ICT (MSIT) in Korea;
Academia Sinica (AS) and National Science and Technology Council (NSTC) in Taiwan. 
This document has been assigned document number LIGO-P2400425.

\appendix

\section{Uniform coordinates for sources in circular binary systems\label{app:binary}}

In this section we work out the uniform-density coordinates $\xi$
for the case of sources in circular  binary systems [Eq.~\eqref{eq:track}]. These choices are motivated
by previous well-established approaches in the literature~\cite{Allen:2002bp,Krishnan:2004sv, 2014PhRvD..90d2002A,
Prix:2006wm, Wette:2018bhc}.

\subsection{Parameter-space resolution}

We start by taking frequency resolution as 
\begin{equation}
    \delta f_{0} = \Tsft^{-1} \,.
\end{equation}
This is a conservative choice~\cite{bretthorstSpectrum,Allen:2002bp,Krishnan:2004sv}. Frequency is
inferred by essentially counting cycles; in that sense higher frequencies may be easier to determine,
but the difference is small enough to be negligible for us for the sake of simplicity.

Similarly, the sky is highly anisotropic, as different Doppler modulations produce different resolutions. We take
the minimal sky resolution to be our sky resolution, in a similar manner to~\cite{Krishnan:2004sv}
\begin{equation}
    \delta \mathrm{sky} = \left(\Tsft v/c\right)^{-2} f_0^{-2}\,.
\end{equation}

For binary orbital parameters,  the general formulae derived following Eq.~\eqref{eq:resolution} tend to 
contain trigonometric expressions such as $\sin\left( \Omega t - \phi_{\mathrm{b}}\right)$ or 
$\cos\left( \Omega t - \phi_{\mathrm{b}}\right)$. We distinguish two cases.
First, the projected semimajor axis $a_{\mathrm{p}}$ and the binary orbital 
phase $\phi_{\mathrm{b}}$:
\begin{equation}
    \frac{\partial}{\partial a_{\mathrm{p}}} f(t;\lambda) = 
    f_0 \Omega \cos\left( \Omega t - \phi_{\mathrm{b}}\right)  \,,
\end{equation}
\begin{equation}
    \frac{\partial}{\partial \phi_{\mathrm{b}}} f(t;\lambda) = 
    f_0 a_{\mathrm{p}} \Omega \sin \left( \Omega t - \phi_{\mathrm{b}}\right)\,.
\end{equation}
To remove the time dependency, we conservatively set trigonometric expressions so that  
the smallest possible resolution is obtained:
\begin{equation}
    \delta a_\mathrm{p} = \Tsft^{-1} \left(f_0 \Omega \right)^{-1} \,,
\end{equation}
\begin{equation}
    \delta \phi_{\mathrm{b}} =  \Tsft^{-1}  \left(f_0 a_{\mathrm{p}}\Omega\right)^{-1} \,.
\end{equation}
Second, the orbital frequency $\Omega$:
\begin{equation}
\begin{aligned}
    \frac{\partial}{\partial \Omega} f(t;\lambda) = \\ 
    f_0 a_{\mathrm{p}}
    \left[
    \Omega t \sin\left( \Omega t - \phi_{\mathrm{b}}\right)
    - \cos\left( \Omega t - \phi_{\mathrm{b}}\right)
    \right]  \,.
    \label{eq:partialOmega}
\end{aligned}
\end{equation}
The sharpest possible resolution is obtained by maximizing Eq.~\eqref{eq:partialOmega}.
We assume to be working in a regime in which the orbital frequency is well resolved,
i.e. $\Omega T_{\mathrm{obs}} \gg 1$. In such case, the sinusoid term dominates and the narrowest resoltuion
is obtained by setting it to 1:
\begin{equation}
    \delta \Omega = \left(\Tsft T_{\mathrm{obs}}\right)^{-1} \left(f_0 a_{\mathrm{p}} \Omega \right)^{-1} \,.
\end{equation}
Note that these resolutions have the same functional dependency on $\lambda$ as those obtained using the
semicoherent metric in the short-segment limit by~\cite{Messenger:2011rg,Leaci:2015bka} (\mbox{$\Tsft \ll 2 \pi / \Omega$} in our notation).

\subsection{Coordinate transformation}

The resulting local density of templates in the parameter space is
\begin{equation}
    \varrho(\lambda) = k f_0^5 a_{\mathrm{p}}^2 \Omega^3 \,,
\end{equation}
with $k = \Tsft^{6} T_{\mathrm{obs}} (v/c)^2$. Searches for this kind of signals are conducted all-sky, in narrow frequency bands,
covering a region of the $(a_{\mathrm{p}}, \Omega)$ plane and the full range
$\phi_{\mathrm{b}} \in [0, 2 \pi)$. The number of templates to place in said region according to our prescription is
\begin{equation}
\begin{aligned}
    \mathcal{N} = &\\
    k
    \int_{f^{\mathrm{min}}}^{f^{\mathrm{max}}} \!\mathrm{d} f_0 f_0^5
    \int_{S^2} \!\mathrm{d} \mathrm{sky}
    \int_{a_{\mathrm{p}}^{\mathrm{min}}}^{a_{\mathrm{p}}^{\mathrm{max}}} \!\mathrm{d} a_{\mathrm{p}} a_{\mathrm{p}}^2
    \int_{\Omega^{\mathrm{min}}}^{\Omega^{\mathrm{max}}} \!\mathrm{d}\Omega \Omega^3 
    \int_{0}^{2 \pi} \!\mathrm{d} \phi_{\mathrm{b}}
    \,.
\end{aligned}
\label{eq:N_binary}
\end{equation}
Two minor details are to be mentioned. First, $\mathrm{d}\mathrm{sky}$ is a solid angle integral.
We conduct our search using equatorial coordinates, which means the corresponding parameters will be 
the right ascension $\alpha$ and the \emph{sine} of the declination $\sin\delta$. Second, this expression explains the usage
of  $\phi_{\mathrm{b}}$ instead of $t_{\mathrm{asc}} = \phi_{\mathrm{b}} / \Omega$, as otherwise the limits in the
$\mathrm{d} t_{\mathrm{asc}}$ would depend on $\Omega$ and the integral would not be separable.

Since the integral in Eq.~\eqref{eq:N_binary} is separable, we use of Eq.~\eqref{eq:xi_of_lambda}
to derive the corresponding uniform coordinates:
\begin{equation}
\begin{aligned}
    \unif{[0]} =& \frac{k^{1/6}}{6} f_0^6 \\
    \unif{[1]} =& k^{1/6} \alpha \\
    \unif{[2]} =& k^{1/6} \sin\delta \\
    \unif{[3]} =& \frac{k^{1/6}}{3} a_{\mathrm{p}}^3 \\
    \unif{[4]} =& \frac{k^{1/6}}{4} \Omega^{4}\\
    \unif{[5]} =& k^{1/6} \phi_{\mathrm{b}}
\end{aligned}\,.
\label{eq:unif_transform}
\end{equation}
Note that overall constants can be freely distributed across all coordinates insofar $\mathrm{d} \unif = \mathrm{d} \lambda \varrho(\lambda)$.

The setup of a template bank is now trivial. First, we transform the parameter-space
limits from $\lambda$ to $\unif$ using Eq.~\eqref{eq:unif_transform}. This results in a hyperbox bounded by
\begin{align}
    \xi_{\mathrm{min}} =& \{\xi(\lambda[k]_{\mathrm{min}}), k = 0, \dots, 5 \}\,,\\
    \xi_{\mathrm{max}} =& \{\xi(\lambda[k]_{\mathrm{max}}), k = 0, \dots, 5 \}\,.
\end{align}
Now, to set up a random template bank, we draw random uniform arrays within the computed bounds 
\mbox{$\xi_i \sim \mathcal{U}(\xi_{\mathrm{min}}, \xi_{\mathrm{max}})$} and map them back to the $\lambda$ space using 
the inverse transformation of Eq.~\eqref{eq:unif_transform} \mbox{$\lambda_i = \lambda(\xi_i)$}.

\bibliography{references}

\end{document}